\documentclass[11pt,prd,aps,onecolumn,showpacs,preprintnumbers,amssymb]{revtex4}
\usepackage{amssymb,epsfig}

\usepackage{amsmath}
\usepackage{graphicx,color}
\usepackage{amssymb,amsthm,subfigure}

\def\Z#1{_{\lower2pt\hbox{$\scriptstyle#1$}}}

\renewcommand{\thefootnote}{\fnsymbol{footnote}}
\newcommand{\nc}{\newcommand}
\nc{\p}{\prime}
\nc{\figcaption}[1]{\def\@captype{figure}\caption{\scriptsize{#1}}}
\nc{\tblcaption}[1]{\def\@captype{table}\caption{\scriptsize{#1}}}

\def\ApJ#1{Astrophys.\ J.\ {\bf#1}}

\begin{document}

\title{Constraining the runaway dilaton and quintessential dark energy}

\author{Ishwaree P. Neupane\footnote{ishwaree.neupane@canterbury.ac.nz}  and
Holly Trowland\footnote{h.trowland@physics.usyd.edu.au}} \affiliation{\vspace{.5cm}\\
Department of Physics and Astronomy, University of Canterbury\\
Private Bag 4800, Christchurch 8020, New Zealand}

\begin{abstract}

Dark Energy is some of the weirdest and most mysterious stuff in
the universe that tends to increase the rate of expansion of the
universe. Two commonly known forms of dark energy are the
cosmological constant, a constant energy density filling space
homogeneously, and scalar fields such as quintessence or moduli
whose energy density can vary with time. We explore one particular
model for dynamic dark energy; quintessence driven by a scalar
dilaton field. We propose an ansatz for the form of the dilaton
field, $|\phi(a)|/m\Z{P} \equiv \alpha\Z1 \ln t+ \alpha\Z2
t^n=\alpha\ln a+ \beta\, a^{2\zeta}$, where $a$ is the scale
factor and $\alpha$ and $\zeta$ are parameters of the model. This
phenomenological ansatz for $\phi$ can be motivated by generic
solutions of a scalar dilaton field in many effective string
theory and string-inspired gravity models in four dimensions.
Using a compilation of current data including type Ia supernovae,
we impose observational constraints on the slope parameters like
$\alpha$ and $\zeta$ and then discuss the relation of our results
to analytical constrains on various cosmological parameters,
including the dark energy equation of state. Sensible constraints
are imposed on model parameters like $\alpha$ and $\zeta$ as well
as on the dark energy/dark matter couplings using results from
structure formation. The constraints of this model are shown to
encompass the cosmological constant limit within $1\sigma$ error
bars.
\end{abstract}

\vskip 1pc 
\preprint{UOC-TP 012/09}

%\begin{titlepage}
\maketitle
%\thispagestyle{empty}
%\end{titlepage}

%\tableofcontents

\renewcommand{\thefootnote}{\arabic{footnote}}
\setcounter{footnote}{0}

\section{Introduction\label{intro}}
%-----------------------------------------------------------------

The discovery of a cosmic acceleration of the
universe~\cite{supernovae,sn2,sn3,sn4} has added a new challenge
for fundamental theories of physics and cosmology. NASA's
observations~\cite{NASA,WMAP1,WMAP2} show that the kind of matter
of which stars and galaxies are made forms less than $5\%$ of the
universe's total mass. Several independent observations indicate
that about $73\%$ of the total energy density of the universe is
in the form of mysterious dark energy or gravitationally repulsive
energy, and about $22\%$ is in the form of non-baryonic cold dark
matter particles which clump gravitationally, but which have never
been directly detected. These scientific enigmas suggest we should
look to new physics beyond the theories of the standard model of
particle physics and Einstein's theory of General Relativity.

The cosmic acceleration of the universe is often attributed to the
existence of some form of dark energy. Understanding the nature of
this dark energy is one of the most challenging theoretical
problems facing present-day cosmology. There is no shortage of
ideas for what dark energy might be, from the quantum vacuum to a
(ultra-) light scalar field, e.g. coming from modified f(R)
gravity~\cite{Carroll:2003,Capozziello:2003gx,Nojiri:2003},
scalar-tensor theories~\cite{Peebles:88A,CW,ZWS,Elizalde:2008},
string-inspired
cosmologies~\cite{Ish05a,CN05,CN05b,Ish06d,TS06,Ish06b}, or
space-time fluctuations of a holographic
origin~\cite{Miao-Li,Wei:2007ty,ISH07a,ISH07b} and the likes. In
some other approaches, cosmic acceleration is a manifestation of
new gravitational physics, which may involve modification of
gravity at long distances~\cite{DDG}, cosmology with extra
dimensions~\cite{Ish03-g,CHNW,Ish03f,Ohta03b}, or warped extra
dimensions~\cite{Ish09a,Ish09b}. There are also proposals which
explain the supernova observations using gravitational effects of
density inhomogeneities~\cite{Wiltshire:2007,DW:07b}. The majority
of work in the subject of cosmic acceleration, as well as the
reviews of dark
energy~\cite{dark-review,Sahni:04,Sahni04c,CST,Sami:09dk}, has
shown that any viable theory of gravity must look in the far
infrared, or on astrophysical scales, like a combination of scalar
fields and gauge fields (weakly) interacting with General
Relativity. Moreover, since the cosmological backgrounds of
interest are time-dependent, for the consistency of any dynamical
dark energy model or theory of cosmic acceleration, such scalar
fields must roll but only slowly as time progresses (see
Ref.~\cite{Guo-Ohta} for some related discussion).

The $\Lambda$CDM model is the simplest known cosmological model
for concurrent universe where it is assumed that most matter is in
the form of cold dark matter, and dark energy is simply described
by the vacuum energy $<T_{\mu\nu}>\Z{\rm vac} \propto \Lambda
g_{\mu\nu}$ in Einstein's field equations
\begin{equation}
R^{\mu\nu}-\frac{1}{2}R g^{\mu\nu} =8{\pi}G \left( T^{\mu\nu}_{\rm
matter} +  T^{\mu\nu}_{\rm vac} \right). \label{einstein field}
\end{equation}
The value of the constant $\Lambda$ is {\it not} set by the
theory~\cite{Weinberg:1988,Sahni02a,Paddy02}, so it may be chosen
to match with observations. Vacuum energy does not vary with space
or time and is not dynamical. This solution seems to be the
best-fit to most observations, especially in regard the dark
energy equation of state, but it does lead to a couple of
extremely unlikely scenarios leading us to question the validity
of using a cosmological constant to describe dark energy. The
first scenario is the ``Why now?" problem. At the present epoch
the universe is composed of approximately $27\%$ matter (both
baryons and dark matter) and $73\%$ dark energy. If we lived in a
slightly earlier or later time these values would be different by
several orders of magnitude, so how come they are comparable at
the current epoch? If the dark energy and matter density were
similar at some earlier epoch, cosmic acceleration would have
began earlier and there would not have been enough time to form
structures, including our galaxy.

It is intuitive to think of the cosmological constant as a vacuum
energy or the energy contained in free space. As the universe
expands, more space is created so there is more gravitational
vacuum energy or free space. As a result the energy density due to
a cosmological constant remains the same throughout time. But,
since a quantum field must be quantized at each and every point in
space, it is also useful to think of a vacuum as not just empty
space but containing a simple harmonic oscillator at each point.
Any vibrations in this field space can propagate through space as
a wave. In the absence of ordinary matter, the hypothetical simple
harmonic oscillators will be in the ground state which has
non-zero vacuum energy. Integrating the ground state energy over
all space and renormalising it gives us the vacuum energy density.
The theoretical value predicted in this way is, however, larger by
a huge factor of at least $10^{15}$ than the mass scale associated
with the observed value of gravitational vacuum energy density,
which is $\rho\Z\Lambda^{1/4}\simeq 10^{-3}~{\rm eV}$. Although
the quantity $\rho\Z\Lambda$ may conceivably be linked with small
numbers such as the neutrino mass ($m_\nu \sim
\rho_\Lambda^{1/4}$), or the ratio $\rho\Z{\rm EW}/\rho\Z{\rm
Planck}\sim \rho\Z\Lambda$, a firm theoretical prediction for the
value of $\rho\Z\Lambda^{1/4}\sim 10^{-3}~{\rm eV}$ is currently
lacking.

The theory confronting a cosmological constant as the source of
dark energy is generally called ``quintessence". In models of
quintessence, one assumes that dark energy is dynamical in nature
instead of being a constant in time and space by utilizing a
standard scalar field. In all modern unified theories including
Kaluza-Klein theory, supergravity and superstring, a fundamental
scalar field called the {\it dilaton} exists along with a spin-2
graviton -- the messenger of gravitational force. In
extra-dimensional theories of gravity, such a scalar field can be
the radion field that encodes the distance between the two branes
(or brane-antibrane pairs) in the fifth dimension. There can be
many such scalar fields in higher dimensional braneworld theories
that may cause cosmic acceleration if they were to self interact
and generate a positive potential. A canonical scalar field $\phi$
having large self interacting potential but almost negligibly
small kinetic term behaves almost as Einstein's cosmological
constant, at least, from a viewpoint of its equation of state.

In addition to exhibiting an effect on the rate of expansion of
the universe, a scalar field quintessence, if it interacts with
dark matter, may lead to an observable effect on how the large
scale structure is growing. Recent observations show that clusters
of galaxies may have formed as a result of small density
fluctuations in the primordial plasma when tiny quantum
fluctuations were amplified by inflation to form pockets of space
that were more dense than the average energy density. Once matter
begins to dominate, gravitational collapse would amplify the tiny
inhomogeneities left by inflation and visible structures begin to
grow into the already formed dark matter halos. In a scenario
where dark energy or a scalar quintessence field interacts with
dark matter, the rate of growth of the dark matter halos would be
affected, and thus the formation of visible structures.

Another important motivation for considering a dynamical scalar
field is an essence of one or more dynamical energy components in
the early universe: so called ``inflaton''. This inflaton field
could easily release its potential energy to matter and radiation
as the field rolls down to its lowest (or metastable) energy
state. It is quite plausible that the late time quintessential
scalar field (together with dilatonic dark matter) is a remnant of
the inflaton field (see, e.g.~\cite{Ish:07a}) although there is no
guarantee they are the same.

In this paper we demand a specific form for the scalar field
itself, which is indeed an approximation to some generic solutions
for a dilaton in some effective string theory models in four
dimensions. Our choice is motivated from the fact that a similar
ansatz would be useful when one is considering the early inflation
of the universe. Starting with the standard form of scalar-tensor
theory of gravity, we will derive field equations that depend on
both the scalar field and the coupling of that field to dark
matter. We impose constraints on the parameters by fitting the
model to observational data from the the cosmic microwave
background (CMB) shift\cite{Komatsu:08a}, baryon acoustic
oscillations (BAO)~\cite{Eisenstein:05a}, SNIa (Gold+HST
sample)~\cite{Riess:06a} and SNLS~\cite{SNLS}. Observational
results from CMB can be used to tightly constrain many useful
cosmological parameters, including the parameters of inflation.

The growth of structure is also expected to depend on the way that
dark energy is coupled to dark matter, granting yet another
constraint. Our focus in this work is to explore ways to constrain
the parameters of the solutions, which include $\chi^2$ curve
fittings to the WMAP+BAO+SNIa+SNLS data sets. Also, we will
consider how the links between quintessential dark energy and
inflation may constrain the parameters of the model. Finally we
discuss on how the results on structure formations constrain the
model parameters and the matter fluctuation growth rate. Using
those constraints we show that the solutions of the model
encompass the cosmological constant limit within $1\sigma$ error
bars.

\section{Quintessence: Theoretical motivation}

Einstein formulated gravity as a tensor theory, where the field
equations of general relativity can be obtained by varying the
action
\begin{equation}\label{einstein-action1}
S= \frac{1}{2} \int \sqrt{-g}\, d^4{x}\, R  + \int \kappa^2 {\cal
L}_m [\Psi_m, g_{\mu\nu}] \sqrt{-g} d^4x,
\end{equation}
where $\kappa{\equiv}\,m_P^{-1}=\sqrt{8{\pi}G}$ is the inverse of
the Planck mass and $G$ is Newton's gravitational constant.
$\Psi_m$ denotes the (ordinary plus) matter. The first integral is
the gravitational part of the action and the second integral is
the matter part, and $\mathcal{L}_m$ is the matter Lagrangian.
Varying this action with the metric $g_{\mu\nu}$ gives Einstein's
field equations
\begin{equation}\label{field2}
\frac{1}{\kappa^2}\left({R_{\mu\nu}-\frac{1}{2}R\,g_{\mu\nu}}\right)
=-\frac{2}{\sqrt{-g}}
\frac{\delta\mathcal{L}_m}{\delta{g^{\mu\nu}}}\equiv{T_{\mu\nu}},
\end{equation}
where $T_{\mu\nu}$ is the energy momentum tensor describing the
density and flux of energy.

Dirac was the first to introduce a scalar field into Einstein
field equations, suggesting that the Newton's constant could be
viewed as a time-dependent parameter. Thereafter Jordan and,
independently, Brans and Dicke introduced a monomial scalar field
known as Brans-Dicke dilaton~\cite{BD}, which also assumes a time
varying gravitational constant and incorporates a scalar field
into Einstein's tensor theory. It was motivated by Mach's
principle, that ``mass there influences inertia here". This was
interpreted by Dicke to mean {\sl the gravitational constant
should be a function of the mass distribution of the universe},
implying that $\frac{1}{G}\propto\frac{M}{R}$, or that a scalar
field somewhat plays the role of the inverse of Newton's constant,
$\frac{1}{G}\propto\varphi$. Unfortunately, the Brans-Dicke
dilaton, with the action
\begin{equation}\label{action-Jordan}
S= \frac{1}{16\pi} \int \sqrt{-g}\, d^4{x} \left( \varphi {R} +
\frac{\omega}{\varphi}\, \partial_i\varphi\,
\partial^i\varphi \right)
 +\int \sqrt{-g}\,
d^4x\, \mathcal{L}_m [g_{\mu\nu}, \Psi_m],
\end{equation}
which is without a field potential for $\varphi$, is unacceptable
since a massless scalar field could easily create a long range
fifth force -- which, however, does not seem to exist in nature.
One can therefore modify the Brans-Dicke type gravitational action
as
\begin{equation}\label{action-Jordan2}
S=\frac{1}{16\pi G} \int \sqrt{-g}\, d^4{x} \left( f(\varphi) {R}+
{\mathcal{L}}_\varphi\right)
 +\int \sqrt{-g}\,
d^4x\, \mathcal{L}_m [g_{\mu\nu}, \Psi_m],
\end{equation}
where ${\mathcal{L}}_\varphi$ is a scalar field Lagrangian in
Jordan frame. In the particular picture of Brane-Dicke model, one
defines $f(\varphi)\propto \varphi $. This is a special case of a
general scalar-tensor theory in the Jordan-Fierz frame, where
$\varphi$ is directly coupled to the Ricci scalar, see, e.g.
Refs.~\cite{Cho:1992,Cho07}.

By using a canonical transformation, we can write the action
(\ref{action-Jordan}) in the Einstein frame
\begin{equation}
S=\int\left(\frac{R}{2\kappa^2}+\mathcal{L}_{\phi}\right)
\sqrt{-g}\,d^4{x}+{\int} \mathcal{L}_m[\widehat{g}_{\mu\nu},
\Psi_m] \sqrt{-\widehat{g}}\,d^4{x}. \label{my action}
\end{equation}
The scalar field Lagrangian ${\cal L}_\phi$ is assumed to have the
standard form
\begin{equation}\label{scalar l}
\mathcal{L}_\phi=-\frac{\gamma}{2}\,g^{\mu\nu}\partial_\mu{\phi}\,
\partial_\nu{\phi}-V({\phi}).
\end{equation}
The Einstein frame metric $\widehat{g}_{\mu\nu}$ is now related to
the Jordan frame metric $g_{\mu\nu}$ via
$\widehat{g}_{\mu\nu}\equiv A(\phi)^2 g_{\mu\nu}$, where $A(\phi)$
is a function of $\phi$. One has $\gamma = \pm 1$ depending on
whether the scalar field is the real or imaginary part of some
complex modulus or axio-dilaton field present in some higher
dimensional theories of gravity. Models similar to the one here
were studied before, see e.g.
Refs.~\cite{Farese:2000ij,Amendola:1999A}, but here we do not
constrain the form of the scalar field potential rather the
evolution of the quintessence scalar field $\phi$. Moreover, we
would assume that $\phi$ is a canonical scalar field ($\gamma=1$)
in Einstein frame and it is coupled to both the ordinary and dark
matter particles but with different gravitational couplings.

To generalize the action (\ref{my action}) further, we assume that
there is a number of different species of matter, including cold
dark matter and baryons -- say $j$ components -- each coupling to
the metric differently with coupling function, $A_j(\phi)$. As the
simplest possibility, we choose exponential couplings
\begin{equation} A_j(\phi) \propto e^{\kappa Q_j\,\phi},
\end{equation}
where, in general, $Q_j$ can be a function of $\phi$. The matter
Lagrangian is now the sum of all components, accounting for all
the different sectors:
\begin{equation}
{\cal L}_m[\tilde{g}_{\mu\nu}, \Psi_m^{(j)}]\equiv \sum_j {\cal
L}_j [e^{2\kappa Q\Z{j}\, \phi} g_{\mu\nu}, \Psi_m^{(j)}].
\end{equation}
We are considering here a scenario in which a purely dark sector
interaction exists, resulting from a nonminimal coupling of dark
matter to a fundamental scalar field or quintessence. Such
couplings give rise to additional forces on dark matter particles
in addition to gravity.

In the early 1990s, Damour, Gibbons and
Gundlach~\cite{Damour:1990} showed that a cosmological model with
two sectors -- cold dark matter with coupling $Q_{c}(\phi)$, and
normal (baryonic) matter with coupling $Q_b(\phi)$ -- satisfies
both the weak equivalence principle (WEP) and constraints from
Solar System observations, provided that
\begin{equation}\label{solar-constr}
m\Z{P} \left\vert\frac{d Q\Z{b}(\phi)}{d\phi} \right\vert
\Z{\phi_0} \lesssim 10^{-2},
\end{equation}
where $\phi\Z0$ is the present-day cosmological background value
of $\phi$. On the other hand, since the nature of both dark energy
and dark matter are still unknown, there is no physical argument
that excludes a possible interaction between them. Furthermore,
the magnitude of the baryonic coupling is constrained from radar
time-delay measurements, $|Q_b|<0.032$ which limits the cold dark
matter coupling to $|Q\Z{CDM}|<1$.

In this paper, we specialize to models where the couplings are
nonnegligible only in the dark sectors, in which case the theory
satisfies the WEP. As long as the scalar field is coupled
non-minimally only to dark matter, the model automatically
satisfies the solar system
constraint~\cite{Damour:1996,Ish07,Neupane07}, see also
Refs.~\cite{Bean:2008ac,Gavela1} for some related discussions.

Let us assume that the spatial curvature of our universe was
erased with inflation in the early universe and choose a spatially
flat Friedmann-Robertson-Walker metric,
\begin{equation}
ds^2=-dt^2+a(t)^2 d\mathbf{x}^2,
\end{equation}
where $a(t)$ is the scale factor, normalized such that $a\equiv
a_0=1$ at present, i.e. at $t=t_0$. This metric assumes the
universe is homogeneous and isotropic on largest scales, which is
a good approximation at large scales and is consistent with CMB
observations~\cite{WMAP1}. For simplicity, we also assume that all
forms of matter, including dark energy, behave as a perfect fluid.
Each form of matter may be characterized by the fluid equation of
state, ${\rm w}_i\equiv {p_i}/{\rho_i}$. In particular, the
cosmological constant may be thought of as a fluid with the
equation of state ${\rm w}_{\Lambda}\equiv
p\Z{\Lambda}/\rho\Z{\Lambda} =-1$. Other forms of matter or energy
are assumed to be perfect fluids as well, for radiation or
relativistic particles, ${\rm w}_\gamma=-1/3$ and for baryons or
cold dark matter ${\rm w}_m\simeq0$. The equation of state for a
scalar quintessence is then defined as
\begin{equation}
{\rm
w}_\phi\equiv\frac{p_\phi}{\rho_\phi}=\frac{\frac{\gamma}{2}\dot{\phi}^2
-V(\phi)}{\frac{\gamma}{2}\dot{\phi}^2+V(\phi)},
\end{equation}
which varies with time.

The variation of the action (\ref{my action}) with respect to
$g^{\mu\nu}$ and $\phi$ gives
\begin{eqnarray}
-\frac{3}{\kappa^2} \, H^2+\frac{\gamma}{2} \, \dot{\phi}^2 +
V\left(\phi\right)+A^4\left(\phi\right)\,\sum_{i}\rho_i=0,
\label{vary1}\\
\frac{1}{\kappa^2}\,\left(2\,\dot{H}+3\,H^2\right)
+\frac{\gamma}{2}\,\dot{\phi}^2-V\left(\phi\right)
+A^4\left(\phi\right)\,\sum_{i}\left({\rm w}_i\,\rho_i\right)=0,
\label{vary2}\\
\gamma\,\left(\ddot{\phi}+3\,H\,\dot{\phi}\right)
+\frac{dV\left(\phi\right)}{d\phi}-
A^3\,\frac{dA\left(\phi\right)}{d\phi}\,\sum_{i}\left(1-3\,{\rm
w}_i\right)\,\rho_i=0, \label{vary3}
\end{eqnarray}
where $H(t)\equiv \dot{a}/a$ is the Hubble parameter. For the
derivation of these equations, see \cite{Ish07,Neupane07,Ish0712}.
In~(\ref{vary3}), it can be seen that $\phi$ couples to the trace
of the stress-energy tensor,
\begin{equation}
T^\mu_{\;\mu}=g^{\mu\nu}T_{\mu\nu}=\sum_{i}\left(1-3\,{\rm
w}_i\right)\,\rho_i.
\end{equation}
For radiation (${\rm w}_\gamma=1/3)$, the stress-energy tensor is
traceless, so there is no interaction between dark energy and
radiation. Equations (\ref{vary1})-(\ref{vary3}) can be
supplemented by a fourth equation arising from the conservation of
energy-momentum tensor, which reads in the Einstein frame as,
\begin{equation}\label{cons}
{\hat{T}}^\mu_{\;(m)\nu ;\,\mu}+{\hat{T}}^\mu_{\;(\phi)\nu
;\,\mu}=0,
\end{equation}
where the semicolon represents a covariant derivative which may be
expanded in terms of the Christoffel symbol as $\hat{T}^\mu_{\;\nu
;\mu}=\hat{T}^\mu_{\;\nu,\mu}+\hat{\Gamma}^\mu_{\;\alpha\mu}
\hat{T}^\alpha_{\;\nu}-\hat{\Gamma}^\alpha_{\;\nu\mu}
\hat{T}^\mu_{\;\alpha}$. An observer unaware of dark energy
looking out at the universe would come to the conclusion that
energy is not conserved. That is not the case, however; it is the
combination of matter and dark energy that is conserved when there
are interactions in cosmology's dark sector. Summing the terms and
remembering that $\hat{T}^\mu_{\;\nu}$ is diagonal and the
relevant Christoffel symbols are $ \hat{\Gamma}^0_{\; 0\,0}=0$,
and $\hat{\Gamma}^1_{\;0\,1}=\hat{\Gamma}^2_{\;0\,2}
=\hat{\Gamma}^3_{\;0\,3}=\hat{H}$ gives the conservation equation
for a perfect fluid
\begin{equation}\label{energy cons}
\dot{\hat\rho}_i+3\,\hat{H}\hat{\rho}_i\left(1+{\rm w}_i\right)=0,
\end{equation}
where $\hat{\rho}_i\propto (a\,A(\phi))^{-3 (1+{\rm w}\Z{i})}$.
The fractional energy densities are defined as
\begin{equation}
\Omega_i=\frac{\rho\Z{i}}{\rho\Z{\rm crit}}=\frac{8\pi
G\rho\Z{i}}{3 H^2},
\end{equation}
so that for each component $0\leq\Omega_i\leq1$. Using the
expressions $\hat{\rho}_i=A^4\rho_i$ and $\hat{a} =A(\phi)a$ and
carefully considering the derivatives involved, the fluid equation
becomes
\begin{equation}\label{vary4}
\dot{\rho_i}+3 H \left(1+{\rm w}_i\right) \rho_i
=\frac{\dot{\phi}}{A(\phi)}\,\frac{dA\left(\phi\right)}{d\phi}\,\left(1-3\,{\rm
w}_i\right)\,\rho_i.
\end{equation}
Out of the four equations, (\ref{vary1})-(\ref{vary3}) and
(\ref{vary4}), only three are independent~\cite{Ish07,Neupane07}.
That is to say, Eq.~(\ref{vary4}) may have been derived from
(\ref{vary1})- (\ref{vary3}), without assuming (\ref{cons}).

From Eqs.~(\ref{vary1}) and (\ref{vary2}), we find that
\begin{equation}
\frac{\ddot{a}}{a}=-\frac{4{\pi}G}{3}
\left[\sum_{i=1}\left(\rho_i+3\,p_i\right)+(\rho\Z{\phi}+3p\Z{\phi})\right].
\label{acceleration}
\end{equation}
A species that contributes to positive cosmic acceleration
($\ddot{a}>0$) must have $(\rho+3p)<0$, i.e. the equation of state
${\rm w}$ must be $<-1/3$. This gives us limits on the range of
values the equation of state which the scalar field may have over
time if it is to contribute a gravitationally repulsive force,
${\rm w}_\phi < -1/3$. Indeed, the WMAP data combined with BAO and
SNIa observations has put much stronger limits on the DE equation
of state, at present ($z\sim 0$), it lies in the range $-1.11<{\rm
w}_{\rm DE} <-0.86\ (95\%\,CL)$ \cite{WMAP2}. This range includes
values of equation of state less than $-1$, which is outside a
theoretical limit set for a canonical scalar field and belongs to
``phantom cosmology''. However, it is possible to get at low
redshifts ${\rm w}_\phi <-1$ by allowing the scalar field to
couple with dark matter non-minimally~\cite{Ish07,Neupane07},
especially by allowing the coupling between $\phi$ and dark matter
to grow with time, or allowing $A(\phi)^2$ to increase
monotonically with $\phi$.

It is practical to express the above equations in terms of the
e-folding time, defined by $N{\equiv}\,\ln[a(t)]$. This is a
useful time parameter when one is considering the expansion
history of the universe, including late time cosmology. We also
utilize the identity $\partial\phi/\partial{N}
=\frac{1}{H}\frac{\partial\phi}{\partial t}$ and denote the
differentiation with respect to $N$ by a prime, $\partial/\partial
N\equiv ^{'}$, and the time derivative represented by a dot,
$\partial/\partial t \equiv {}^\cdot$. Using following definitions
\begin{eqnarray}
&&  \epsilon\equiv\frac{\dot{H}}{H^2}, \quad
Q\equiv\frac{d\ln\left[A\left(\phi\right)\right]}{d(\kappa\phi)},
\quad \Omega_i\equiv\frac{\kappa^2\,A^4\,\rho_i}{3\,H^2}, \quad
\Omega_{\phi}\equiv \frac{\kappa^2\,\rho_{\phi}}{3\,H^2}, \quad
{\rm w}_{\phi}\equiv \frac{p_{\phi}}{\rho_{\phi}}, \label{subs1}
\end{eqnarray}
one can write the equations (\ref{vary1})-(\ref{vary3}) and
(\ref{vary4}) in the following form
\begin{subequations}
\begin{align}
\sum_{i}\Omega_i+\Omega_{\phi}=1,
\label{vary6}\\
2\,\epsilon+3\,\left(1+{\rm
w}_{\phi}\right)\,\Omega_{\phi}+3\,\sum_{i}\left(1+{\rm
w}_i\right)\,\Omega_i=0,
\label{vary7}\\
\Omega_{\phi}^\prime+2\,\epsilon\,\Omega_{\phi}+3\,\Omega_{\phi}\,
\left(1+{\rm w}_{\phi}\right)+\phi^\prime\,Q\,\sum_{i}(1-3\,{\rm
w}_i)\,\Omega_i=0,
\label{vary8}\\
\sum_{i}\Omega_i^\prime+2\,\epsilon\,\sum_{i}\Omega_i +3\,\sum_{i}
\Omega_i\,\left(1+{\rm
w}_i\right)-\phi^\prime\,Q\,\sum_{i}(1-3\,{\rm w}_i)\,\Omega_i=0,
\label{vary9}
\end{align}
\end{subequations}
where the sum over $i$ represents the sum over all forms of matter
or energy. It is important to note that in minimal coupling case,
$A(\phi)=1$, $Q$ vanishes and the field equations have one less
parameter making them easier to handle. Solutions of these
equations can be used to reconstruct the behavior of $\phi$ over
time. The behavior of the scalar field is generally characterized
by its potential, which is given by
\begin{equation}\label{potentiala}
\kappa^2 V\left(\phi\right) = H^2(\phi)\,\left( 3\Omega_{\phi}
-\frac{\gamma\kappa^2} {2}\,{\phi^\prime}^2\right).
\end{equation}
For simplicity, one can make a specific ansatz for the form of the
potential in order to reduce the number of degrees of freedom and
hence solve the set of field equations analytically. For many
potentials discussed in the literature, the initial value of
$\phi$ and $\dot{\phi}$ must be finely tuned to obtain the correct
values of $\Omega\Z\phi$ and ${\rm w}_{\phi}$ today. The tuning of
the initial field expectation value is required in addition to
tuning the potential parameters.

Given the many alternative form of (quintessential) potentials it
is useful to try and understand the properties of DE in a
model-independent manner. In this paper, we will make an ansatz
for the form of the scalar field so that the form of the
potential, together with a number of cosmological parameters, can
be reconstructed. This is a valid way to study the effects of the
scalar field on the background since it is required that the
potential must be fairly flat today, or, equivalently, satisfy
that $(\dot{\phi}/m_{Pl}) \ll \sqrt{3} H$. This is because the
scalar field must be rolling slowly (with a small kinetic term) to
force the equation of state close to $-1$ at $z\gtrsim 0$, as
inferred by cosmological observations (the WMAP data combined with
BAO and SNIa observations).

% From (\ref{vary7}), we get
% \begin{equation}\label{equstate-DE}
% {\rm w}_{\phi}=-\frac{2\,\epsilon+3\,\sum_{i} \left(1+{\rm
% w}_i\right)\,\Omega_i+3\,\Omega_{\phi}}{3\,\Omega_{\phi}},
% \end{equation}
An unambiguous expression for the quintessence equation of state
(EoS) is given by
\begin{equation}\label{assumption4}
 {\rm w}\Z{\phi}(a)=\frac{\gamma \kappa^2
\left(\phi^\prime\right)^2-3\, \Omega_{\phi}}{3\,\Omega_{\phi}}.
\end{equation}
This result is obtained by substituting the expression for
$V(\phi)$, Eq.~(\ref{potentiala}), into the definition of ${\rm
w}_\phi$ in Eq.~(\ref{subs1}). In this note, rather than
parametrizing the dark energy equation of state ${\rm w}_\phi$
with two or more phenomenological parameters, we will choose to
parametrize the evolution of the field $\phi$ (cf see
Eq.~(\ref{ansatz})). This, in turn, will induce a dependence of
${\rm w}_\phi$ on those parameters through
Eq.~(\ref{assumption4}). The same applies to the scalar field
potential $V(\phi)$ and the Hubble parameter which both depend on
these parameters through Eqs.~(\ref{potentiala}) and
~(\ref{hubble}). In our analysis, we only consider a canonical
quintessence with $\gamma= 1$, and use the relation
(\ref{assumption4}) which is an exact and definitive expression
for the equation of state.

Other parameters of interest are the effective equation of state
and the deceleration parameter. The effective equation of state
describes the average equation of state, taking into account all
the energy components of the universe:
\begin{equation}\label{effective1}
{\rm w}_{\rm eff} \equiv \frac{p_{\rm tot}}{\rho_{\rm tot}}, \quad
p_{\rm tot}\equiv p_{\phi}+ \sum_{i} p_i \, e^{4\kappa Q_i\phi},
\quad \rho_{tot} \equiv \rho_{\phi}+ \sum_{i} \rho_i \,e^{4 \kappa
Q_i \phi}.
\end{equation}
From Eq.~(\ref{vary7}) and the definitions in (\ref{subs1}), we
find that
\begin{equation}\label{ptot}
p_{\rm tot}=\frac{3\,H^2(\phi)}{\kappa^2}\,
\left(-\frac{2\,\epsilon}{3}-\Omega_{\phi}-\sum_{i}\Omega_i\right),
\quad {\rm w}_{\rm eff}=-1-\frac{2\,\epsilon}{3}.
\end{equation}
Another important quantity is the deceleration parameter
\begin{equation}\label{q}
q\equiv -(1 +\epsilon) =
-\frac{\ddot{a}}{a\,H^2}=\frac{d}{dt}\left(\frac{1}{H}\right)-1.
\end{equation}
This parameter is defined such that it is negative for an
accelerating expansion ($\ddot{a}>0$) and positive for a
decelerating expansion ($\ddot{a}<0$).

\section{Motivation and Analytic solutions} \label{solns}

In order to be able to analytically solve the generalized field
equations obtained in Sec. II, there must be more assumptions made
to reduce the number of degrees of freedom. In this paper, instead
of making an ansatz for the scalar field potential or a
paramterization of the dark energy equation of state, we wish to
reconstruct relevant cosmological parameters from some
phenomenologically well-motivated Ans\"atze for a scalar field
quintessence. Our approach is new in the existing literature.

In the literature, basically, there are two simple and convenient
approximations to the form of a time-evolving scalar field
quintessence, see, e.g.
Refs.~\cite{ART93,Gasperini:01a,Barreiro:98a}. Those
approximations are \begin{equation}\label{ansatz-1st-form} \kappa
|\phi| = \alpha\Z1 \ln t+ {\rm const}, \quad \kappa {|\phi|}=
\alpha\Z2 t^{m}+ {\rm const},
\end{equation}
These evolutions for $\phi$ generally represent the tracking
limits of some more general solutions for a runway dilaton in many
string-inspired scalar-tensor
theories~\cite{Ish06b,ART93,Gasperini:01a}. Similar approximations
have been adopted also in a few phenomenologically motivated
models for quintessence, see, e.g.
Refs.~\cite{Chevallier:00,Linder02a,Barreiro:98a,
Picon:00a,Picon01a,Scherrer:2007a,Scherrer08b,Dutta08a}.

In this paper, we will assume that the time-evolution of
quintessence scalar field $\phi$ is well described by a
combination of the above two specific solutions. Using a general
feature of the scale factor, that $a(t)\propto t^{p}$, with
different values of $p$ at different epochs, such as $p=2/3$ in a
matter-dominated epoch and $p>1$ in a dark-energy-dominated epoch,
the time-evolution of $\phi$ may be written as
\begin{equation} \label{ansatz}
\kappa\left(\phi\Z{0}-\phi(a)\right) = \alpha \ln a + \beta\,
a^{2\zeta},
\end{equation}
where $\alpha$, $\beta$ and $\zeta$ are some new arbitrary
constants, and $a\equiv a(t)$ is the scale factor of a
four-dimensional FRW universe. This parameterization of $\phi$,
which might bear some generic features of a fundamental scalar
dilaton field or metric moduli in some string-inspired models, can
also be motivated from other two aspects. First, it really gives a
useful information as regard the dark energy equation of state
once the parameters like $\alpha$ and $\zeta$, or their
combination, are known (even approximately) using observational
data. Second, it certainly helps to explain the cosmic coincidence
problem, since the value of an arbitrary coefficient $V_0$ that
arises with {\it a prior} choice of the scalar field potentials,
such as $V(\phi)=V_0\,e^{-\lambda (\phi/m_P)}$, $V(\phi)=V_0
\phi^{-n}$ and $V(\phi)=V_0\left(1- c_0 \phi^2\right) e^{-\lambda
(\phi/M_P)}$, won't be very important in our approach. Indeed, the
parameterization~(\ref{ansatz}) for the evolution of $\phi$ is
neither more arbitrary nor more restrictive than other
parametrizations and approaches to quintessential dark energy in
the literature. See
Refs.~\cite{Sahni:2006pa,Scherrer:2007a,Scherrer08b,
Dutta08a,Samushia:2008fk,Ratra08a} for some other plausible ways
of reconstructing dark energy or quintessential potentials.

Our ansatz for $\phi$, i.e.~(\ref{ansatz}), has nonetheless some
similarities with respect to numerous dark energy potentials
proposed previously, which may be reconstructed by using
Eq.~(\ref{potentiala}), or alternatively, $V(\phi)=\kappa^2
\left[(3+\epsilon)(1-\Omega_m)+\frac{1}{2} \Omega_m^\prime\right]
H^2(\phi)$. Especially, with $\zeta\simeq 0$ in (\ref{ansatz}), we
find that the leading term of the reconstructed potential is
simple exponential in $\phi$,
\begin{equation}
V(\phi)\equiv V_0\, e^{\,\alpha (\phi/M_P)}+ \cdots,
\end{equation}
where the dots represent some other terms which could arise, for
instance, due to the effects of matter-scalar interactions.
Similarly, with $\alpha \simeq 0$, the scalar potential takes the
form
\begin{equation}
V(\varphi) \equiv V\Z{1}\, e^{-\,\zeta\varphi^2} \left(1-
c\Z{0}\varphi^2\right)+\cdots,\label{quin-pot2}
\end{equation}
where $\varphi \equiv  |\Delta\phi|/m\Z{P} =
(\phi\Z0-\phi)/m\Z{P}$ and $c\Z{0}=\frac{2\zeta^2}{3}$. A
polynomial potential multiplied with exponential pre factor as
above may be motivated by string theory and standard Kaluza-Klein
gravity. An effective dark energy potential as above was proposed
in~\cite{Albrecht99}. In our approach, the total or effective
quintessence potential is roughly given by a linear combination of
the above two potentials. Interestingly, in our discussions below,
the actual form of the potential is not very important, but only
the values of the slope parameters $\alpha$ and $\zeta$.

In fact, using the freedom to rescale $N$ ~($\equiv \ln a$) or
shift $\phi$, we can set $\beta=1$~\cite{Ish0712}. The model then
contains two free parameters, $\alpha$ and $\zeta$, so there is
more freedom to tune them according to the observational
constraints. In our model, there is one more degree of freedom
that must be constrained. This extra parameter, which is the
coupling constant between the dark matter and the scalar field, is
a mixed blessing -- it makes the system easier to tune but it also
makes the solutions more complicated.

In order to completely solve Eqs.~(\ref{vary6})-(\ref{vary9}), one
must specify some initial conditions. These conditions can be
defined on the current composition of the universe, as measured by
WMAP plus other observations~\cite{WMAP2}, and are $\Omega_b\simeq
0.05$, $\Omega_{CDM}\simeq 0.22$, $\Omega_{\rm r}\simeq10^{-4}$
and $\Omega_\phi\simeq 0.73$. The baryon component (subscript $b$)
is quite small compared to the cold dark matter component
(subscript $CDM$), so even though there are tight limits on any
baryon-dark-energy coupling, it is sufficient to assume a general
dark energy coupling to $\Omega_m$,
($\Omega_m\equiv\Omega_b+\Omega_{CDM}\simeq0.27$). Henceforth the
only components of the universe that are not considered negligible
are dark matter and the dark energy component, up to $z\sim {\cal
O} ({10}^2)$, when radiation starts playing a major role. Also, it
is assumed that the dark matter equation of state is the same as
that of ordinary matter (or dust), ${\rm w}_m\simeq 0$. Using
these assumptions, and for simplicity setting $\kappa=1$ from now
on, the system of equations (\ref{vary6}-\ref{vary9}) is
simplified to
\begin{subequations}
\begin{align}
\Omega_m+\Omega_{\phi}=1
\label{vary10}\\
 2\,\epsilon+3\,\left(1+{\rm w}_{\phi}\right)\,\Omega_{\phi}+3\,\Omega_m=0
\label{vary11}\\
\Omega_{\phi}^\prime+2\,\epsilon\,\Omega_{\phi}+3\,\Omega_{\phi}\,
\left(1+{\rm w}_{\phi}\right)+\phi^\prime\,Q\,\Omega_m=0
\label{vary12}\\
\Omega_m^\prime+2\,\epsilon\,\Omega_m+3\,\Omega_m-\phi^\prime\,Q\,\Omega_m=0.
\label{vary13}
\end{align}
\end{subequations}
Equations (\ref{vary10})-(\ref{vary13}) can now be solved to find
explicit expressions for ${\rm w}_\phi$, $\epsilon$ and
$\Omega_m$. Solving the equations directly gives the solution for
$\Omega_m$ and $\epsilon$ in terms of $\alpha$, $\zeta$ and $Q$
($\alpha$ and $\zeta$ are parameters from the ansatz for $\phi$,
and $Q$ is the coupling term).

The explicit solutions for $\Omega_m$ and $\epsilon$ are given by
\begin{equation}\label{sol-Om-int}
\Omega_m=\frac{X(N)}{C\Z0 -3\int X(N)\, dN },\quad
\epsilon=-\frac{3}{2} \Omega_m -\frac{1}{2} {\phi^\prime}^2=
-\frac{3}{2}\Omega_m -\frac{1}{2}\left(\alpha+2\zeta e^{2\zeta
N}\right)^2,
\end{equation}
where
\begin{equation}
X=X(N=\ln a) \equiv \exp\left[-(3-\alpha Q-\alpha^2)N+(2\alpha+Q)
e^{2\zeta N}+ \zeta e^{4\zeta N}\right].\label{soln-X}
\end{equation}
The scale factor $a(t)$ is normalized such that $a(z=0)\equiv
a\Z0=1$, so in the past $a < 1$ and hence $N\equiv \ln a < 0$. The
relationships among the scale factor $a$, redshift $z$ and
e-folding time $N$ are
\begin{equation}
\frac{a}{a_0}=\frac{1}{1+z}, \quad N=\ln(a/a_0)=-\ln(1+z).
\end{equation}
So, at $t=t\Z0$, $z=N=0$. Without loss of generality, we set
$a_0=1$, so $N=\ln a$. To solve the integral (\ref{sol-Om-int}),
in the small $\zeta$ limit, we may use the approximation $e^{\zeta
\ln a}\approx 1+ \zeta\ln a$. The solution for $\Omega_m$ is now
given by
\begin{equation}\label{main-soln-Omm}
\Omega_m=\frac{3-\tilde{\alpha}^2-Q \tilde{\alpha}} {3+C_1
\exp{\left[\left(3-\tilde{\alpha}^2 -Q \tilde{\alpha}\right)N
\right]}},
\end{equation}
where we have introduced a new variable $\tilde{\alpha}$ such that
\begin{equation}
\tilde{\alpha}\equiv\alpha+2\zeta.
\end{equation}
The solution for $\Omega_m$ automatically gives the solution for
$\Omega_\phi$ which is simply $\Omega_\phi=1-\Omega_m$, in using
Friedmann constraint. The integration constant $C_1$ can be fixed
by using the initial conditions of the model, assuming
$\Omega_m(t_0)\equiv \Omega\Z{m 0}$, where $t_0$ is the time now.

Particularly, at low redshifts, the terms quadratic (and higher
powers) in ${\zeta N}$ contribute only subdominantly. [Even at
high redshifts, the terms like $(2\alpha+Q)e^{2\zeta N}$ and
$\zeta e^{4\zeta N}$ in (\ref{soln-X}) are only sub-leading to the
first term in $X(N)$, i.e. $-(3-\alpha^2-\alpha Q) N$, since
$\zeta>0$ (by assumption) and $N\equiv \ln {a}< 0$ in the past.]
To quantify this, one can write
\begin{equation}
e^{\zeta \ln a}= 1+ \zeta \ln a + \frac{\zeta^2
\ln{a}^2}{2}+\cdots = 1+\zeta\ln a + \zeta^2 |\ln a|+\cdots.
\end{equation}
For instance, in between the redshifts $z=2$ and $z=0$, $N\equiv
\ln {a}$ runs from $-1.09$ to $0$, and the solution
(\ref{main-soln-Omm}) remains valid for $\zeta$ as large as $\zeta
\sim 1/2$, but in the discussions below we will always assume
rather implicitly that $\zeta \ll 1/2$.

The Hubble parameter may be evaluated by solving the differential
equation associated with the relation $H\epsilon =H'$. This again
introduces an integration constant which can be fixed by the
normalization $H(t_0)\equiv H_0$. From the solutions for
$\Omega_m$ and $\epsilon$, the other parameters discussed in Sec.
II can also be derived. The expression for ${\rm w}_\phi$ is given
by rearanging Eq.~(\ref{vary11}), while $V(\phi)$ is given by
Eq.~(\ref{potentiala}) using the solutions for $\Omega_m$ and $H$.

From the second equation in (\ref{sol-Om-int}), we can see that
(retaining the 4d gravitational constant $\kappa$)
\begin{equation}
\kappa^2 {\phi^\prime}^2 = 1+ q -\frac{3}{2} \Omega\Z{m}.
\end{equation}
Requiring that $q<0$ (as implied by the type Ia supernovae) and
imposing a generous lower bound on the value of $\Omega_m$, which
is $\Omega_{0 m} >0.24$, one obtains the safe upper bound
$$\kappa |\phi\Z{0}^\prime | < 0.8. $$
However, to make the present model compatible with various other
data sets, including WMAP observations, one may be required to
satisfy $\kappa |\phi\Z{0}^\prime | \lesssim 0.4$, which
encompasses the cosmological constant limit within $1\sigma$
error.

\section{Constraints on the model}\label{constraints}

\subsection{Constraints from supernova}\label{sn1a}

In this subsection, we use several sets of data from recent
cosmological observations and put constraints on our model,
limiting the values of $\alpha$, $\zeta$ and $Q$. To do this, code
for $\chi^2$ curve fitting given by Nesseris and Perivolaropoulos
\cite{limits} has been utilized along with our solution for ${\rm
w}_\phi$ and the standard form for the Hubble parameter, in terms
of the two parameters
\begin{equation}\label{definew0}
p_1\equiv\tilde{\alpha}^2 , \quad p_2\equiv-Q\tilde{\alpha},
\end{equation}
where, as before, $\tilde{\alpha}\equiv\alpha+2\zeta$. We use the
Hubble parameter to fit our model to data on the expansion rate of
the universe. The Hubble parameter can be found from the solution
for $\epsilon$ and $\Omega_m$ and the solution to the differential
equation, $\epsilon=\dot{H}/H^2$, which is the definition of
$\epsilon$. After some simplification, we find that the Hubble
parameter is given by
\begin{equation}\label{hubble}
H=H\Z0\left(\Omega_{0 m}\,a^{-3-p_2}+(1-\Omega_{0
m})\,a^{-p_1}\right)^{1/2}.
\end{equation}
This result is obtained by integrating out the expression
$\epsilon\equiv H^\prime/H =\left(\ln H\right)^\prime$ (see,
Eq.~(\ref{sol-Om-int})) and substituting the expression of
$\Omega_m$ from Eq.~(\ref{main-soln-Omm}). The frame where matter
density decreases as $\rho_m\propto 1/(A(\phi) a(t))^3$ (where the
scale factor is modified by the coupling $A(\phi)$) is known as
the Einstein frame. In our expression for the Hubble parameter
(\ref{hubble}) the matter part is $\Omega_{0\,m}a^{-3-p_2}$, where
the density evolution is modified by the parameter $p_2$ which is
associated with the coupling $Q$. Thus this standard form of the
Hubble parameter is in the Einstein frame.

\begin{figure}[!ht]
%\FIGURE{\label{best-fit1} \vbox{\vskip 10 pt
\centerline{\includegraphics[width=3in,height=3.2in]{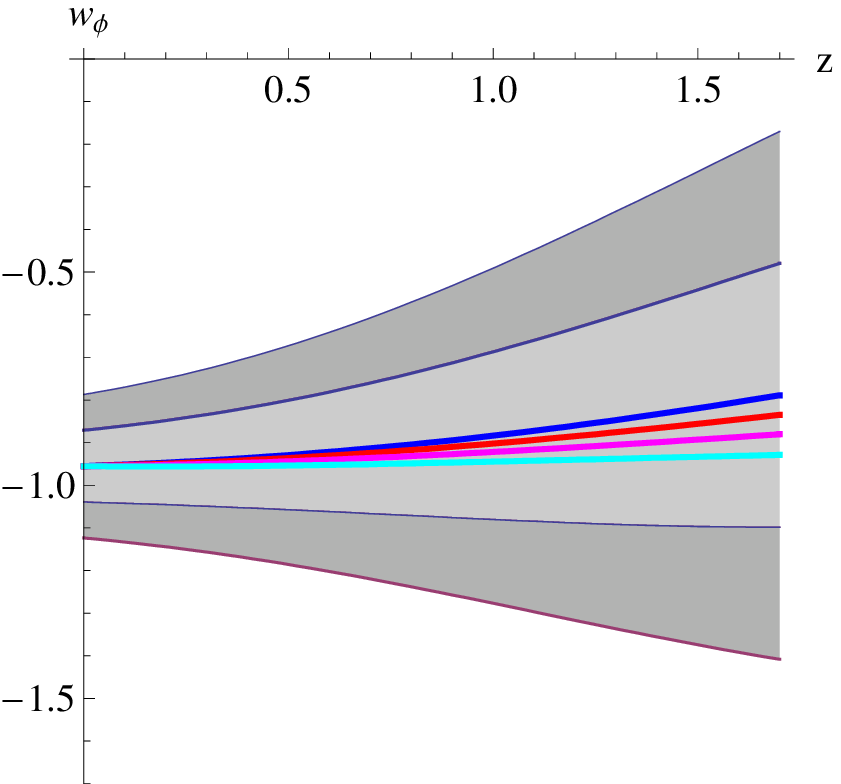}\hskip0.2in
\includegraphics[width=3in,height=3.2in]{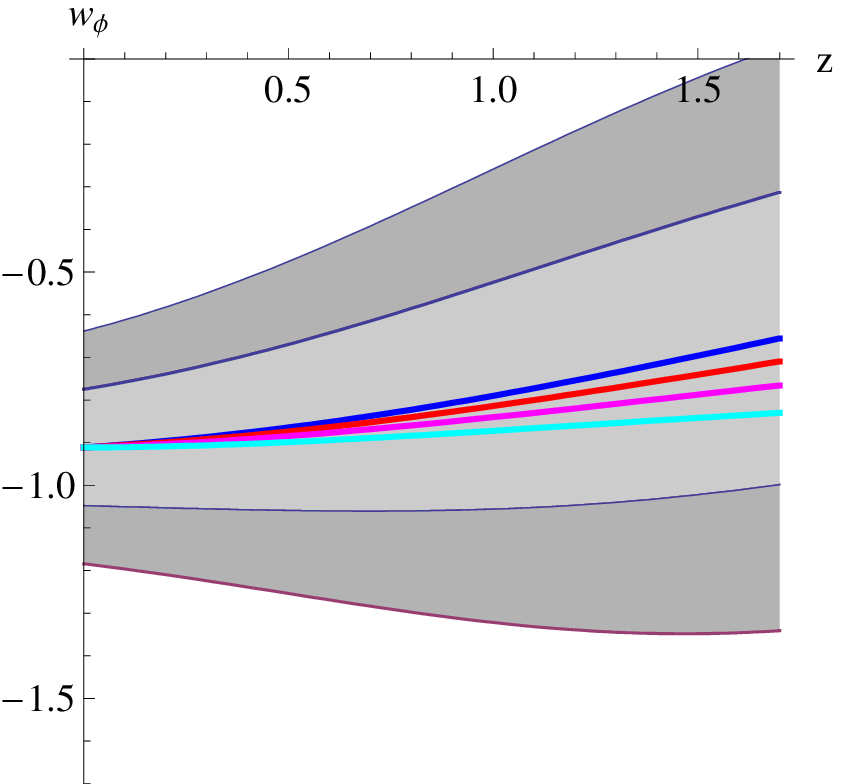}}
\caption{Scalar field equation of state ${\rm w}_\phi$ as a
function of redshift ($z$) with $1\sigma$ and $2\sigma$ errors
(light and dark gray shades) and with EoS solutions given by
$\zeta=0.2,\, 0.15,\, 0.1,\, 0$ (top to bottom; green, pink, red,
blue online), $\Omega_{0_m}=0.27$. \textbf{Left plot}:
SN1a+WMAP+BAO+SNLS. \textbf{Right plot}: SN1a only
}\label{best-fit1}
\end{figure}

As in some standard approaches, let us first drop the scalar field
- dark matter coupling. The model then reduces to one-parameter
parameterization of the Hubble expansion rate. Remember that the
model still deviates from the $\Lambda$CDM model, since $p_1\ne 0$
even though $p_2=0$, leading to a time-varying equation of state
for dark energy (see cf Fig.~\ref{best-fit1}). Needless to say,
the $\Lambda$CDM cosmology corresponds to the choice
$\tilde{\alpha}=0=Q$. This can easily be understood from the
following observation. With $\kappa \phi^\prime \simeq
\alpha+2\zeta \equiv \tilde{\alpha}$, we get
\begin{equation}\label{main-wphi}
{\rm
w}_\phi(z)=\frac{{\tilde{\alpha}}^2-3\Omega_\phi}{3\Omega_\phi}
=\frac{(\tilde{\alpha}^2-3)
C\Z{1}-3\tilde{\alpha}Q\,(1+z)^{3-\tilde{\alpha}^2-\tilde{\alpha}Q}}{3
C\Z{1}+ 3\tilde{\alpha}(\tilde{\alpha}+Q)
(1+z)^{3-\tilde{\alpha}^2-\tilde{\alpha}Q}}.
\end{equation}
$C\Z{1}$ in the above equation or in (\ref{main-soln-Omm}) is
fixed such that $\Omega_m(t_0)=\Omega_{m0}$ at $z=0$. Note that
the dark energy EoS ${\rm w}_\phi(z)$ is time-varying as long as
$\tilde{\alpha}\ne 0$.

{One notes that {\it a prior} choice or an arbitrary
parameterization of ${\rm w}(z)$ could easily lead to an erroneous
reconstruction of the dark energy equation of state. To see this,
one considers the parameterization $H^2(z)=H_0^2 \left[\Omega_{0m}
(1+z)^3 +\Omega_{\rm DE}\right]$, where $\Omega_{\rm DE}\equiv
(1-\Omega_{0m})\exp\left\{3\int\frac{1+{\rm
w}(z)}{1+z}\,dz\right\}$. For example, with the ansatz $w(z)\equiv
{\rm w}_0+ \frac{z}{1+z}\,{\rm w}_1$, one finds that $\Omega_{\rm
DE}= (1-\Omega_{0m})\exp\left\{\frac{3 {\rm w}\Z{1}}{1+z}+(1+{\rm
w}_0+{\rm w}\Z{1})\ln (1+z)-{\rm w}_1+{\rm const} \right\}$. Now,
with ${\rm w}_1=0$, one has $H^2(z)=H_0^2 \left[\Omega_{0m}
(1+z)^3 +(1-\Omega_{0m})\times {\rm const}\times (1+z)^{3(1+{\rm
w}_0)}\right]$, which has a form similar to Eq.~(\ref{hubble}),
especially, with $p_2=0$, i.e. $Q=0$. This could give a wrong
impression that in our model the choice $Q=0$ gives a constant
equation of state for dark energy. The dark energy equation of
state is not constant as long as $\tilde{\alpha}\ne 0$, or
precisely, when $\alpha, \zeta\ne 0$}.

In Table 1 we present the best-fit values using only one parameter
(i.e. $p_1\ne 0$ and $p_2=0$) first using only the Gold sample of
157 type Ia supernova data, the Supernova Legacy Survey (SNLS)
data alone and then for combined data sets. The combined data
includes the cosmic microwave background (WMAP) shift, baryon
acoustic oscillations (BAO), suvernovae type Ia Gold sample (SNIa)
and legacy survey (SNLS). The errors of these fits are shown in
Fig.~\ref{best-fit1}. For the combined data sets $\chi^2$ is
minimum when $\Omega_{0m}\approx 0.27$ where
$\alpha+2\zeta\simeq0.3\pm 0.3$. For $\Omega_{0\,m}>0.29$ in the
fit to the Gold sample alone, the EoS drops below -1, indicating
phantom quintessence
 and an imaginary $\tilde{\alpha}$.
% Later in Table 3, we present results when both $p_1$
% and $p_2$ take non-zero values.
\vspace{0.3cm}
\begin{table}
Table 1: The best fit of $\tilde{\alpha}$ to expansion history
data, $Q=0$:
\begin{tabular}{|l|l|l|l|l|}
\hline $\Omega\Z{0 m}$& \quad \qquad $p_1$ & \qquad
$|\tilde{\alpha}| $
& \quad $w\Z{\phi}(z=0)$ (mean) & \quad $\chi\Z{min}^2$ \\
\hline
\multicolumn{5}{l}{SNIa Gold data sets}\\
\hline
 $ 0.26$ & \quad $0.271\pm 0.286$ & \quad $0.521\,(\pm 0.275)$ & \quad $-0.878$ & \quad $177.98$ \\
 $0.27$ & \quad $0.194\pm 0.298$  & \quad $0.441\,(\pm 0.339)$ & \quad $-0.911$ & \quad $177.76$ \\
 $0.28$ & \quad $0.112\pm 0.312$  & \quad $0.335\,(\pm 0.466)$ & \quad $-0.948 $ & \quad $177.54$  \\
%$0.30$ & $-0.065\pm 0.344$  & $ - $ & $  -1.03 $ & $177.09$ \\
\hline
\multicolumn{5}{l}{SNLS data sets}\\
\hline
$0.26$ & \quad $-0.006\pm 0.274$ & \quad $-$ & \quad $-1.00$ & \quad $104.15$ \\
$0.27$ & \quad $-0.081\pm 0.285$ & \quad $-$ & \quad $-1.04$ & \quad $104.13$ \\
$0.28$ & \quad $-0.159\pm 0.297$  & \quad $-$ & \quad $-1.07 $ & \quad $104.12$  \\
%$0.30$ & $-0.329\pm 0.323$  & $ - $ & $ -1.16 $ & $104.12$ \\
\hline
\multicolumn{5}{l}{WMAP+BAO+SNIa+SNLS data sets}\\
\hline
$0.26$ &  \quad $0.123\pm 0.178$  & \quad $0.350\pm 0.254 $ & \quad $-0.945$ & \quad $283.54$ \\
$0.27$ &  \quad $0.098\pm 0.1842$  & \quad $0.313\pm 0.294 $ & \quad $-0.955$ & \quad $283.24$ \\
$0.28$ & \quad $ 0.072\pm 0.190$  & \quad $0.268\pm 0.354 $ & \quad $-0.967 $ & \quad $283.61$  \\
%$0.30$ & $0.018\pm 0.203$   & $0.133\pm 0.762 $ & $-0.991$ & $284.19$ \\
\hline
\end{tabular}
\end{table}
\medskip

It is indeed the SNLS data that lowers $w\Z{\phi}(z)$ towards the
value $-1$ at present, i.e. at $z=0$, which is clearly seen from
the best-fit values in Table 1.
%\begin{subequations}
%\begin{align}
%\Omega_{0 m}=0.26 : \quad p_1=-0.006 \pm 0.274, \quad
%\chi^2=104.14,\\
%\Omega_{0 m}=0.27 : \quad p_1=-0.081 \pm 0.285, \quad
%\chi^2=104.13,\\
%\Omega_{0 m}=0.28 : \quad p_1=-0.159 \pm 0.297, \quad
%\chi^2=104.12,
%\end{align}
%\end{subequations}
The SNLS data naively suggests a small cross over range between
the cosmological constant ${\rm w}\Z\Lambda=-1$ and the phantom
dark energy ${\rm w}\Z{\phi}<-1$ (or, equivalently, $p_1 <0$), but
the error bars are too large. For the combined data sets (from
WMAP, BAO, SNIa and SNLS) the best-fit value of ${\rm w}_\phi$
falls in the range $-0.94
> {\rm w}_\phi >-1$.

\begin{figure}[!ht]
%\FIGURE{\label{best-fit2}\vbox{\vskip 10 pt
\centerline{\includegraphics[width=3in,height=3.2in]{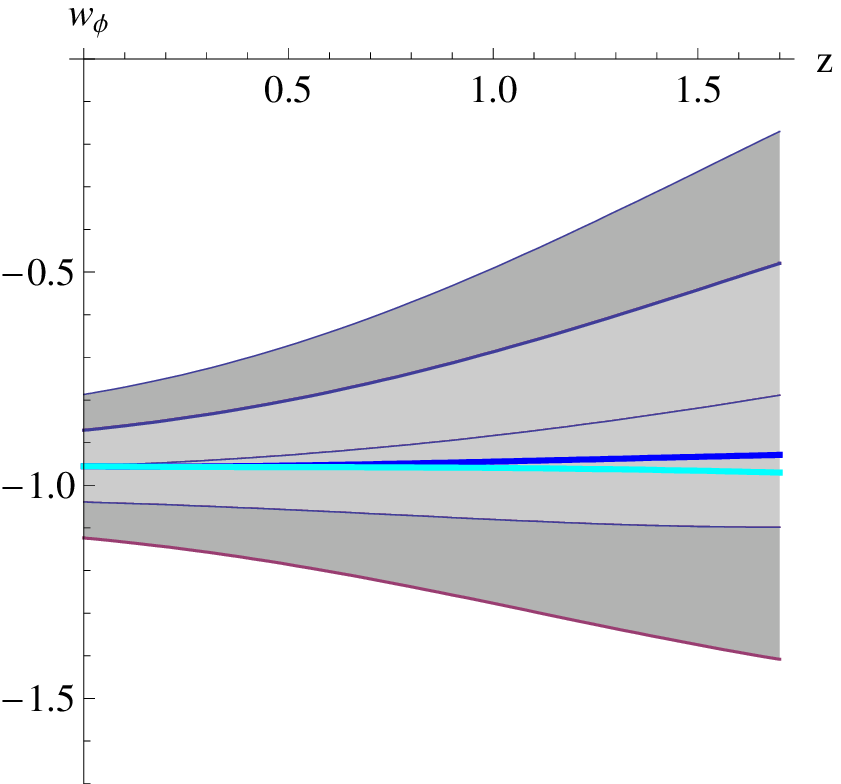}\hskip0.2in
\includegraphics[width=3in,height=3.2in]{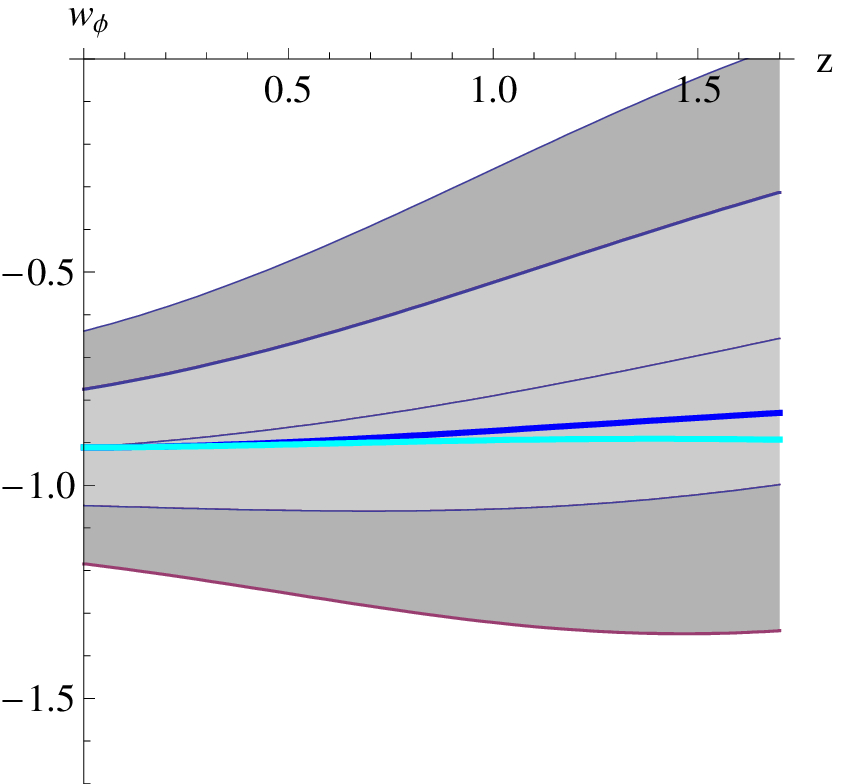}}
\caption{As in Fig.~\ref{best-fit1} but with $\zeta=-0.2$ (lower
line, cyan) and $\zeta=+0.2$ (upper line, blue). \textbf{Left
plot}: SN1a+WMAP+BAO+SNLS. \textbf{Right plot}: SN1a only}
\label{best-fit2}
\end{figure}

From the relationships above, and also noticing that for the
combined data sets the scalar field equation of state is closer to
$-1$ over a longer time in Fig.~\ref{best-fit1}, we find that the
solution using all the data sets is compatible with a cosmological
constant. This analysis has provided a limit to the relationship
between $\alpha$ and $\zeta$ but there is still a degree of
freedom for choosing the value of $\alpha$ or $\zeta$. This degree
of freedom proves hard to constrain since for other observational
tests, it is only required that $\alpha$ and $\zeta$ be of
comparable magnitude, or that $\zeta=0$. The plots of ${\rm
w}_\phi$ in Figure \ref{best-fit1} show that the chi-squared is
minimized when $\zeta$ is small since cosmological reconstructions
for large $\zeta$ diverge from the best-fit line.
Fig.\ref{best-fit2} displays the same constraints on the model,
this time showing the effect of positive or negative $\zeta$.
There is more divergence from the best-fit line for negative
values of $\zeta$, suggesting that positive $\zeta$ is a better
fit.

\vspace{0.3cm}

\begin{figure}[!ht]
%\FIGURE{\vbox{\vskip 10 pt
\centerline{\includegraphics[width=2.9in,height=2.5in]{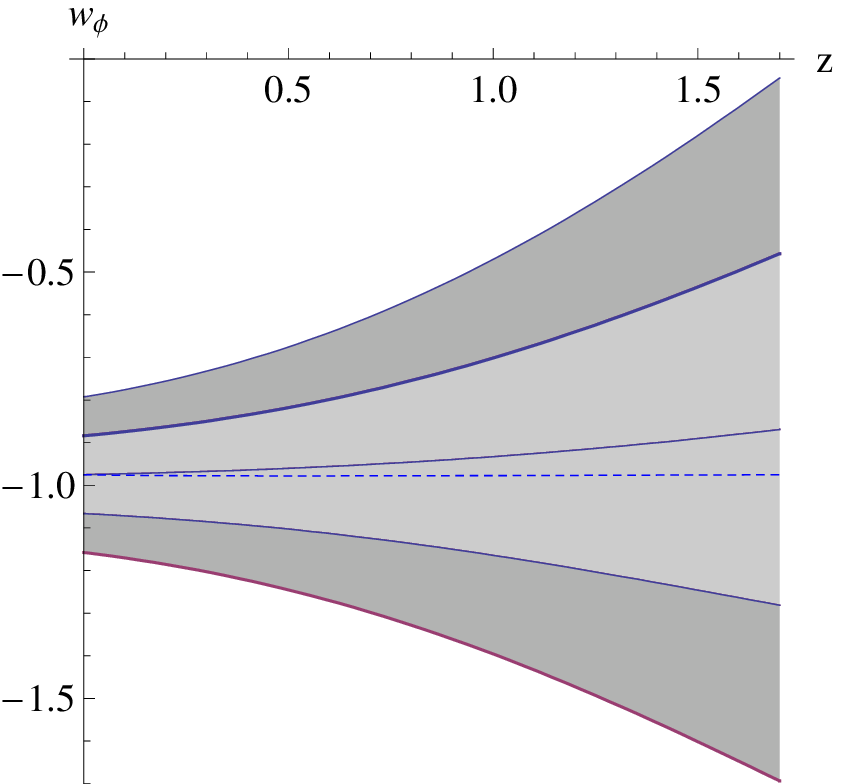}\hskip0.2in
\includegraphics[width=2.9in,height=2.5in]{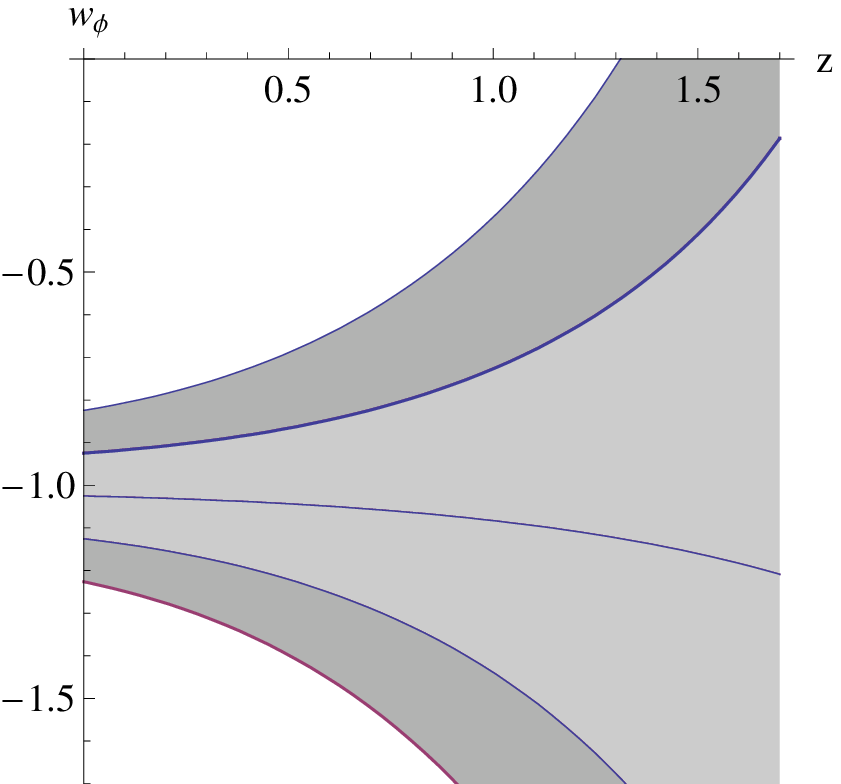}}
\caption{Best-fit plot of ${\rm w}_\phi$ with $Q\ne 0$ (i.e.
$p_2\ne 0$). The continuous line is the best fit and the dashed
line is for $\zeta=0.2$.
 The inner and outer
shaded regions represent $1\sigma$ and $2\sigma$ error bars.
\textbf{Left plot}: $\Omega_{0m}=0.26$. \textbf{Right plot}:
$\Omega_{0m}=0.28$.} \label{new-best-fit1}
\end{figure}

\begin{table}
Table 2: The best fit values for both $\tilde{\alpha}$ and $Q$
(WMAP+BAO+SNIa+SNLS):
\begin{tabular}{|c|c|c|c|c|c|c|}
\hline $\Omega\Z{0 m}$& $p_1$ & $p_2$ & $|\tilde{\alpha}| $ & $Q$
& $w\Z{\phi}(z=0)$& $\chi\Z{min}^2$ \\
\hline
$0.26$ & $0.055\pm0.203$ & $0.015\pm0.021$ & $0.235\pm0.432$ & $-0.065\pm0.149$ & $-0.98$& $ 283.02$\\
$0.27$ & $0.001\pm0.210$ & $0.021\pm0.021$ & $0.027\pm3.83$ & $-0.772\pm 1.08$ & $-1.00$ & $282.22$\\
$0.28$ & $-0.055\pm0.217$ & $0.027\pm0.021$ &-&-& $-1.03$ &$281.98$\\
$0.29$ & $-0.111\pm0.225$ & $0.032\pm0.021$  &-&-& $-1.05$ &$282.24$\\
$0.30$ & $-0.168\pm0.233$ & $0.037\pm0.021$ &-&-& $-1.08$ &$283.00$\\
\hline
\end{tabular}
\end{table}
\medskip

Our proposed model has the freedom of allowing for a non-zero
coupling of dark matter to dark energy. Including this in the fit
to the data sets gives the results of Table 2. Note that for
 $\Omega_0>0.27$ we get negative values of $p_1$. From the definition
  of $p_1$, this means an imaginary $\tilde{\alpha}$ and $Q$, and a
  negative equation of state which is phantom quintessence. The best-fit
  to the data is for $\Omega_{m0}\simeq 0.275$ which is
apparently the best fit value for $\Lambda$CDM model obtained from
the combined WMAP5+BAO+SNIa datasets~\cite{WMAP2}. The best fit
value for $\tilde{\alpha}$ is close to zero, the cosmological
constant limit.
 $Q$ is constrained to $-0.772\pm 1.08$ but this error is too
 large for this to be a conclusive result. It is clear from the shaded
  error region in Fig. \ref{new-best-fit1} that adding the extra
   parameter, $Q$, greatly increases the uncertainty in the fit.

\subsection{Constraints from structure formation}\label{structure}

Cosmological models are probed by observing the effects of dark
energy on the expansion of the universe through measuring the
distances to far off galaxies~\cite{Cattoen:08a}. Other than this
there are not many ways to check models that account for the late
time acceleration of the universe. However, if dark energy does
interact with dark matter even weakly, it may have an observable
effect on the early stages of structure formation.

Since dark matter naturally guides the way for the formation of
observed structure, any interaction between dark energy and dark
matter would have an effect on the manner that visible structure
was formed. In particular, we investigate how the rate of
structure growth is affected by a non-zero coupling of dark matter
to a fundamental scalar field.

Matter fluctuations evolve according to the standard linear
differential equation
\begin{equation}\label{fluctuation}
\ddot{\delta}+2H\dot{\delta}=4\,\pi G \rho_m \delta,
\end{equation}
where $\delta\equiv\delta\rho_m/\rho_m$ is the linear matter
density contrast. This linear growth equation comes from the
perturbed equations of motion of Einstein's general relativity,
see, e.g. Ref.~\cite{weinberg}. It may be numerically solved to
reconstruct the growth of these matter fluctuations, but as yet
there is no physical theory that relates the matter density
contrast to the matter density. In \cite{Lahav:1991} it was
proposed that the growth rate of matter perturbations, defined as
\begin{equation}
f\equiv\frac{\delta^\prime}{\delta}=\frac{1}{\delta}
\frac{d\delta}{d\ln a}=\frac{1}{\delta} \frac{d\delta}{dN},
\end{equation}
can be characterized by the following simple expression
\begin{equation}\label{fom}
f=\left[\Omega_m(N)\right]^\eta.
\end{equation}
This ansatz works well at low redshifts $(z\lesssim 2$) and for
coupled dark energy models with a small coupling parameter. For
the $\Lambda$CDM model, $\eta\simeq 0.56$ \cite{Lahav:1991}, and
for modified DGP gravity, $\eta\simeq 0.68$ \cite{Guzzo:2008}.
However, this ansatz does have the drawback that $f\leq 1$, since
$0\leq\Omega_m\leq1$. This is fine for the standard $\Lambda$CDM
model which assumes $f=1$ during the matter-dominated era for high
redshift, but does not hold for all models of dark energy.

For the present model of quintessential dark energy, as discussed
in~\cite{Amendola:2003a} in more detail, the linear growth
equation (\ref{fluctuation}) is modified to be
\begin{subequations}
\begin{align}
\delta_c^{\prime\prime}+\left(2+\varepsilon+Q_c\,\phi^\prime\right)
\delta_c^\prime =\frac{3}{2}\left(1+2 Q_c^2\right)\delta_c
\Omega_c + \frac{3}{2}(1+2Q_b Q_c)\delta_b
\Omega_b,\label{modi-fluc1}\\
\delta_b^{\prime\prime}+\left(2+\varepsilon+Q_b\,\phi^\prime\right)\delta_b^\prime
=\frac{3}{2}\left(1+2 Q_b^2\right)\delta_b \Omega_b +
\frac{3}{2}(1+2Q_b Q_c)\delta_c \Omega_c,\label{modi-fluc2}
\end{align}
\end{subequations}
for baryonic and CDM components respectively. The couplings
constant $Q_b$ and $Q_c$ are usually coupled, so in general it is
not possible to express these equations as a single differential
equation. Amendola~\cite{Amendola:2003a} made a naive estimation
that in the limit $|Q_b|\ll |Q_c|$ and $\Omega_b \ll \Omega_c$,
one can write
\begin{equation}\label{eqn-amendola}
\delta^{\prime\prime}+ (2+\epsilon+Q \,\phi^\prime)\delta^\prime
=\frac{3}{2}\Omega_m (1+2 Q^2)\delta,
\end{equation} where $Q\equiv Q_c$ and $\delta \approx
\delta_c$ (to leading order). In a sense, the baryonic component
$\Omega_b$ is assumed to be negligible. But one should also note
that $\delta_b$ and its derivatives are non-negligible, otherwise
Eq.~(\ref{modi-fluc2}) would be inconsistent. Equation
(\ref{eqn-amendola}), which reduces to the standard result when
$Q=0$, is a close approximation rather than being an exact result.
In the following discussion, we will assume that $\delta_c
\Omega_c \gg \delta_b \Omega_b$.

In space-time backgrounds dominated by baryonic matter, the effect
of dark matter may be neglected ($\Omega_c=0$). The effective
Newton's constant $\widehat{G}$, and two dimensionless post
Newtonian parameters $\hat{\gamma}$ and $\hat{\beta}$ are related
to the coupling constants
\begin{equation}
Q(\phi)\equiv m_P \frac{\partial \ln A(\phi)}{\partial \phi},
\quad X(\phi) \equiv m_P \frac{\partial Q(\phi)}{\partial\phi}
\end{equation}
via~\cite{Damour:1993}
\begin{equation}
\widehat{G}= G \left [ A^2(\phi) (1+ Q^2)\right]_{\phi_0}, \quad
\hat{\gamma}=1- \frac{2 Q^2}{1+ Q^2}, \quad \hat{\beta}=
1+\frac{Q^2 X}{2(1+Q^2)^2}.
\end{equation}
For an exponential coupling $A(\phi)\propto e^{Q_b\kappa \phi}$,
with $Q_b$ behaving (almost) as a constant, the local gravity
constraint $|1-\hat{\gamma}|< 10^{-4}$~\cite{Bertotti:2003}
implies that $Q_b^2 \equiv (1-\hat{\gamma})/(1+\hat{\gamma})<
10^{-4}$ and hence $|Q_b|<0.01$. This constraint is still weaker
than the one arising from weak equivalence principle
violation~\cite{Damour:2002,Chiba:06a}. With $Q_b\simeq $ const,
one has $|1-\hat{\beta}| \simeq 0$ to a large accuracy. The
effective Newton's constant is now modified but it must be within
the limit for the time variation of Newton's constant
$|\frac{d{\widehat{G}}/dt}{G}| <10^{-14}{\rm yrs}^{-1}$, see e.g.
Refs.~\cite{limits,Nesseris07}, which is generally the case when
$|Q_b| \lesssim 0.01$ is satisfied. In most of the discussion
below we will assume that $|Q_C|\gg |Q_b|\simeq 0$ and $Q_c \equiv
Q$.

In the nonminimally coupled theory, the linear growth rate
(\ref{fluctuation}) is modified from the uncoupled case; the
quantity $\delta\equiv\delta\rho_m/\rho_m$ now depends on the
values of $Q$ and $\phi^\prime$. The result of this is a
modification of the expression for the growth rate,
Eq.~(\ref{fom}), where it is renormalized by either the parameter
$\eta\rightarrow\tilde{\eta}$ and/or a coefficient $f_0$:
\begin{equation}
f=f_0\left[\Omega_m(N)\right]^{\tilde{\eta}}.
\end{equation}
Following the analysis in Ref.~\cite{Di-Porto07}, if the growth
rate is modified by a coefficient, it may be dependent on the
coupling by
\begin{equation}\label{approx-sol}
f=\left[\Omega_m(N)\right]^\eta \left(1+c\,Q^2 +\cdots \right),
\end{equation}
which reduces to (\ref{fom}) for minimal coupling ($Q=0$). This
expression of $f$ is merely a phenomenological fit where $\eta$
and $c$ may be determined by fitting the standard approximate
solution (\ref{approx-sol}) to a numerical solution to
Eq.~(\ref{modi-fluc1}) or Eq.~(\ref{eqn-amendola}). The best-fit
parameters found from this numerical analysis in
Ref.~\cite{Di-Porto07} are $\eta\simeq 0.56$ and $c\simeq 2.1$.
The physical theory pertaining to the dependence of $\Omega_m$ and
$Q$ on the perturbation growth rate is as yet unknown so this kind
of generalized fit is just an approximation. Determining the
growth rate over time and comparing to observational results gives
limits to the model that are independent of limits from fitting
the model to supernova and WMAP data sets. This grants an
independent check for the consistency of the model.

To preform this check of our model, we use some known observed
values for the growth rate at different low redshifts (listed
below in table), from which we can determine the values of $f_0$
and $\eta$ that best match the data. There are very few
measurements of the growth rate at low redshifts and often $f$ can
be evaluated by assuming a $\Lambda$CDM model, for which
$\eta\approx 0.56$.

\begin{table}
Table 3: Observed matter fluctuation growth rate as compiled
in~\cite{Wei:2008,Wei08b,Alam:2008at}.
\begin{tabular}{|c|c|c|c|c|c|c|}
\hline $z$ &  $ 0.15$  & $0.35 $ & $0.55 $ & $0.77 $ & $1.4$ & $3$
\\ \hline
$f_{\rm obs} $ &  $0.51\pm 0.11$  & $0.70\pm 0.18 $
& $0.75\pm 0.18$ & $0.91\pm 0.36 $ & $0.90 \pm 0.24$ & $1.46\pm 0.29$ \\
\hline
\end{tabular}
\end{table}

\begin{figure}[!ht]
%\FIGURE{\vbox{\vskip 10 pt
\centerline{\includegraphics[width=3.0in,height=1.8in]{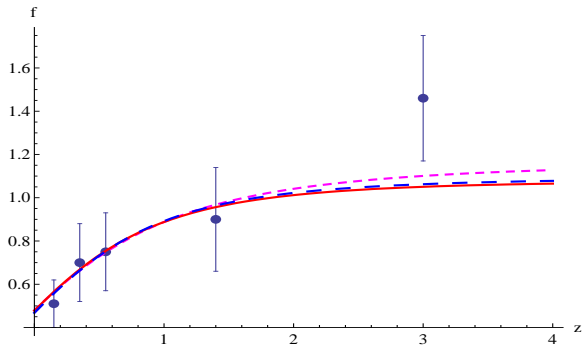}}
 \caption{Best fit curve to the observed values of the
growth rate, with $\Omega_{0m}=0.27$. The red solid line
corresponds to a best fit of Q, $\tilde{\alpha}$ and $\eta$, the
blue dashed line corresponds to $\eta=0.56$ with best fit for Q
and $\tilde{\alpha}$, and the blue dotted line corresponds to
$\eta=0.68$ with best fit for Q and $\tilde{\alpha}$.}
\label{structurefit}
\end{figure}

It is clear from Eqs.~(\ref{main-soln-Omm}) that $\Omega_m$ is in
fact a function of $\tilde\alpha$ and $Q$ and thus our
parameterization of $f$, i.e. (\ref{approx-sol}), is implicitly a
function of $\tilde\alpha$. The fits are illustrated by
Fig.~\ref{structurefit}, where the best-fit parameters are
\begin{equation} |Q|=0.199, \quad \tilde{\alpha} = -0.010, \quad \eta=0.538.
\end{equation}
In the above we have used the normalization $f_0=1+2.1Q^2$ used
in~\cite{Di-Porto07}. Within the $1\sigma$ errors, the
normalization constant $f_0=1.16\pm 0.34$ is compatible with a
minimally coupled quintessence scenario. The least-squares curve
fitting of all three of $\tilde{\alpha}$, $Q$ and $\eta$ to the
data often leads to huge ranges of possible values of these
parameters. For brevity, we may restrict ourselves to fitting only
two parameters such as $Q$ and $c$. The best-fit value of the
normalization constant $c$ is quite sensitive to the choice of
$\eta$, or vice versa. For example, assuming $\Omega_{m0}=0.27$,
$\tilde{\alpha}=0$ and $\eta=0.56$ and fitting for $Q$ and $c$
gives
\begin{equation}
Q=0.42 \pm 0.10, \qquad c=1.17 \pm 0.02.
\end{equation}
Similarly, holding $\eta=0.68$ and $c=2.1$ give the values
\begin{equation}
\quad |Q|=0.41\pm0.14, \qquad \tilde{\alpha}=0.57\pm0.38.
\end{equation}
In the above analysis we have used an explicit analytical
expression for $\Omega_{m}$, which is nothing but the
equation~(\ref{main-soln-Omm}). If it is instead the parameter
$\eta$ that is renormalized, say $\eta\simeq
0.68$~\cite{Guzzo:2008} as in modified gravity proposed by Dvali,
Gabadadze and Porrati, the fit to the data in
Fig.~\ref{structurefit} seems to be closer.

% One may find the best-fit values of $Q$ and $A$ using some other
% inputs. For example, taking $\alpha+2\zeta \simeq 0$ and making
% the ansatz
% $$ f(z)=(1+2.1 Q^2) \,\left(\Omega_m (z)\right)^A, $$
% we find that $(Q, A)=(0.19, 0.57)$ and $(0.18, 0.63)$ respectively
% for $\Omega_{0m}=0.24$ and $\Omega_{0m}=0.27$. In obtaining these
% values we have dropped the last entry from table 3 (as the results
% from high redshift data are not quite unanimous).

\subsection{Constraints from CMB}

\begin{figure}[!ht]
%\FIGURE{\vbox{\vskip 10 pt
\centerline{\includegraphics[width=3.1in,height=2.2in]{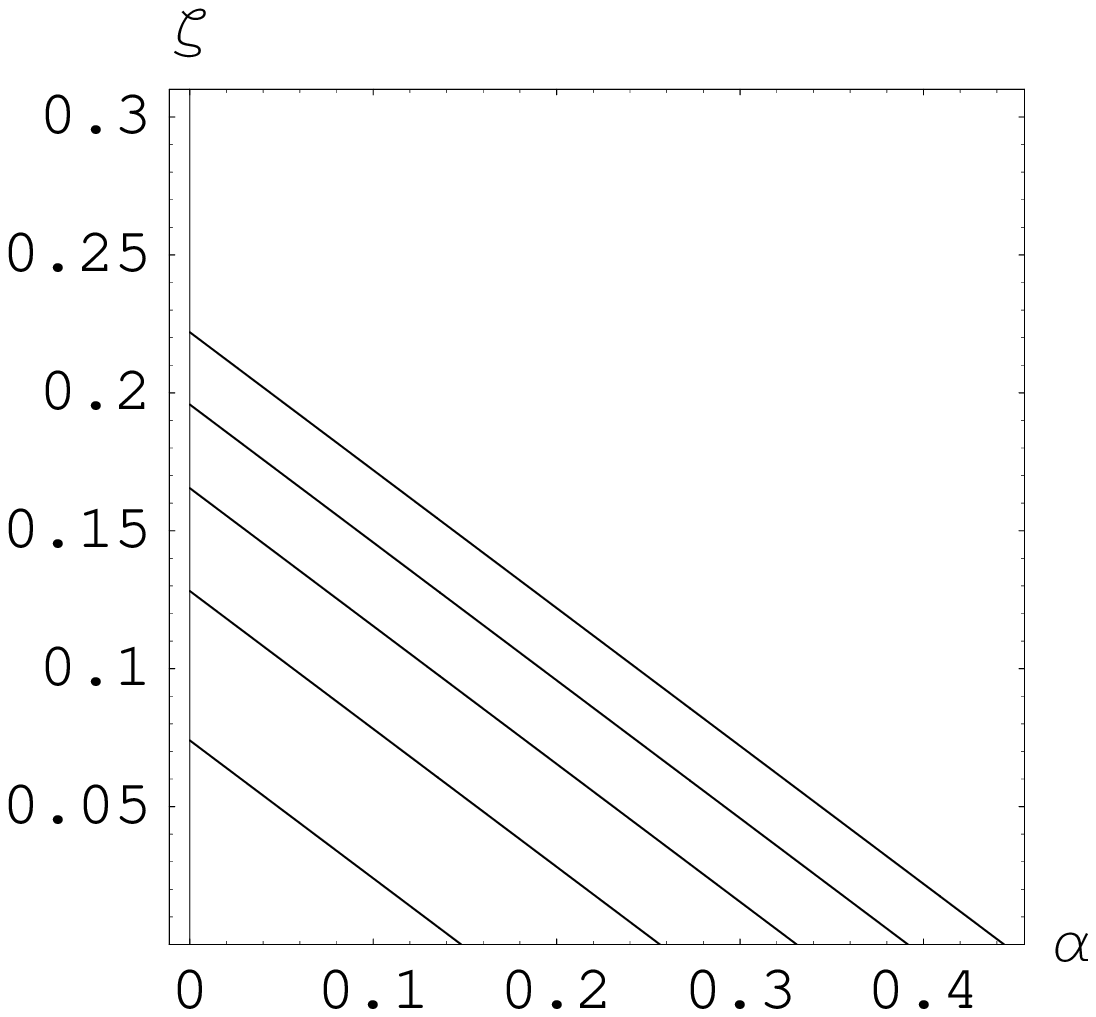}\hskip0.1in
\includegraphics[width=3.1in,height=2.2in]{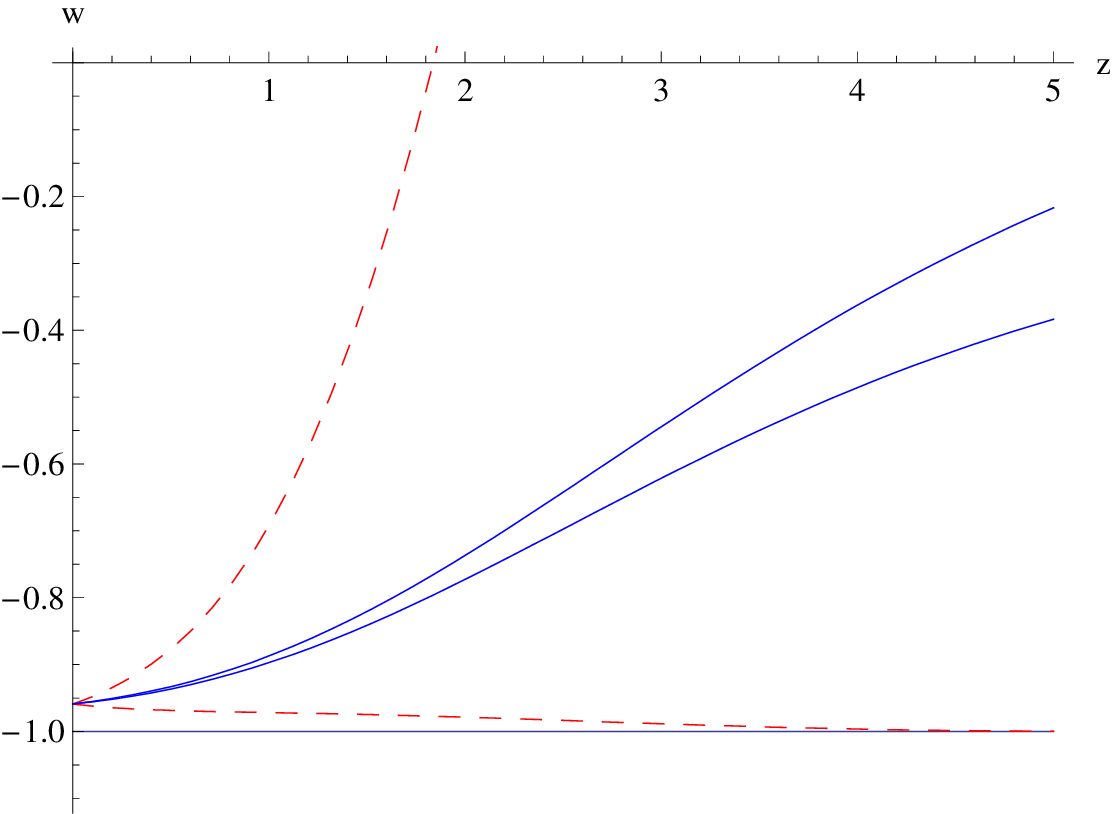}}
\caption{\textbf{Left plot} Contour plot of ${\rm w}_\phi$ in the
range $(-0.91,-0.99)$ (top to bottom) at $z=0$. \textbf{Right
plot} The dark energy EoS for the two solutions of $\zeta=\pm0.05$
(solid lines) and $\zeta=\pm0.25$ (dashed lines).}\label{contour}
\end{figure}

Another test of our model to the observational constraints is to
demand that the equation of state presently lies in the range,
$-1.11<{\rm w}_\phi<-0.86$, as implied by the WMAP data combined
with BAO and type Ia SN~\cite{WMAP2}. This is illustrated in the
contour plot, Fig.~\ref{contour}, where the outer contour
corresponds to ${\rm w}_\phi=-0.91$ and the innermost contour
corresponds to ${\rm w}_\phi=-0.99 $. For ${\rm w}_\phi$ to lie
within the WMAP5 limit $-1.11<{\rm w}_\phi<-0.86$, the constraint
is
\begin{equation}\label{cmbconstraint}
\tilde{\alpha}=0(\pm 0.5)
\end{equation}
where the value of the coupling has little impact. This
relationship holds for small $|\zeta|\lesssim 0.1$. It is
reasonable to assume that the value of the scalar field equation
of state has not changed much in the recent past. Using this
tighter restriction, by inspecting Fig.~\ref{contour}, we get a
tighter relationship between $\alpha$ and $\zeta$, as
$z\rightarrow 3$, error in relationship (\ref{cmbconstraint})
goes to $\pm 0.2$. % $\alpha\rightarrow-2 \zeta\;(\pm 0.2)$

There is still a degree of freedom in the choice of the value of
$\alpha$ or $\zeta$, so we look to a possible link between
inflation and late time acceleration to set it. Quintessence is
designed such that the scalar field that drives the late time
acceleration may be the same, or an evolved form of the inflaton
field that drove inflation. If we assume that it is in fact the
same field that drives both of these accelerating epochs, we may
impose extra conditions on the model by requiring the quintessence
field to satisfy the constraints of inflation. One such
constraining parameter is the spectral index, which describes the
slope of the angular power spectrum of the CMB. The WMAP data
inferred a red-tilted spectrum, $n_s<1$, which is consistent with
most inflationary models. The spectral index can be approximated
in the small, positive $\zeta$ limit, as~\cite{Ish:07a}
\begin{equation}\label{ns}
n_s\simeq1-\alpha^2.
\end{equation}
For the WMAP+BAO+SN mean value of $n_s=0.960\,^{+0.014}_{-0.013}$
\cite{WMAP2} and $|\alpha|=0.20\,\pm0.04$. By using the
relationship between $\alpha$ and $\zeta$ determined by the
combined data sets in Table 2 ($|\alpha+2\zeta|\simeq0.3$), we
obtain $\zeta=\pm0.05\pm0.15$ or $\zeta=\pm0.25\pm0.15$. The
evolution of the EoS for these solutions of $\zeta$ are plotted
 in Figure \ref{contour}. The EoS for $\zeta=\pm0.25$ deviates far from -1 for a positive value
 of $\alpha$. Both signs of $\alpha$ are possible so the lower value $\zeta=0.05$ may be more
 physical.

\section{Dynamical behaviour}

In this section we examine how the scalar field varies over time
and how this depends on the parameters $\alpha$, $\zeta$ and $Q$.
As an example we choose the value $\alpha=0.2$ and using
$\alpha+2\zeta\simeq 0.3$ from Table 1, $\zeta=0.05$. From
structure formation we estimate $|Q|=0.2$. In our analysis perhaps
the most relevant parameter is the field velocity of $\phi$, i.e.
$\phi^\prime \equiv d\phi/d\ln a$, which is a measurable quantity,
at least in principle, either by accurately measuring the equation
of state ${\rm w} \Z{\phi}(z)$, or by placing a constraint on the
dark matter-scalar coupling $Q$, or both.

\begin{figure}[!ht]
%\FIGURE{\vbox{\vskip 10 pt
\centerline{\includegraphics[width=3.0in,height=2.0in]{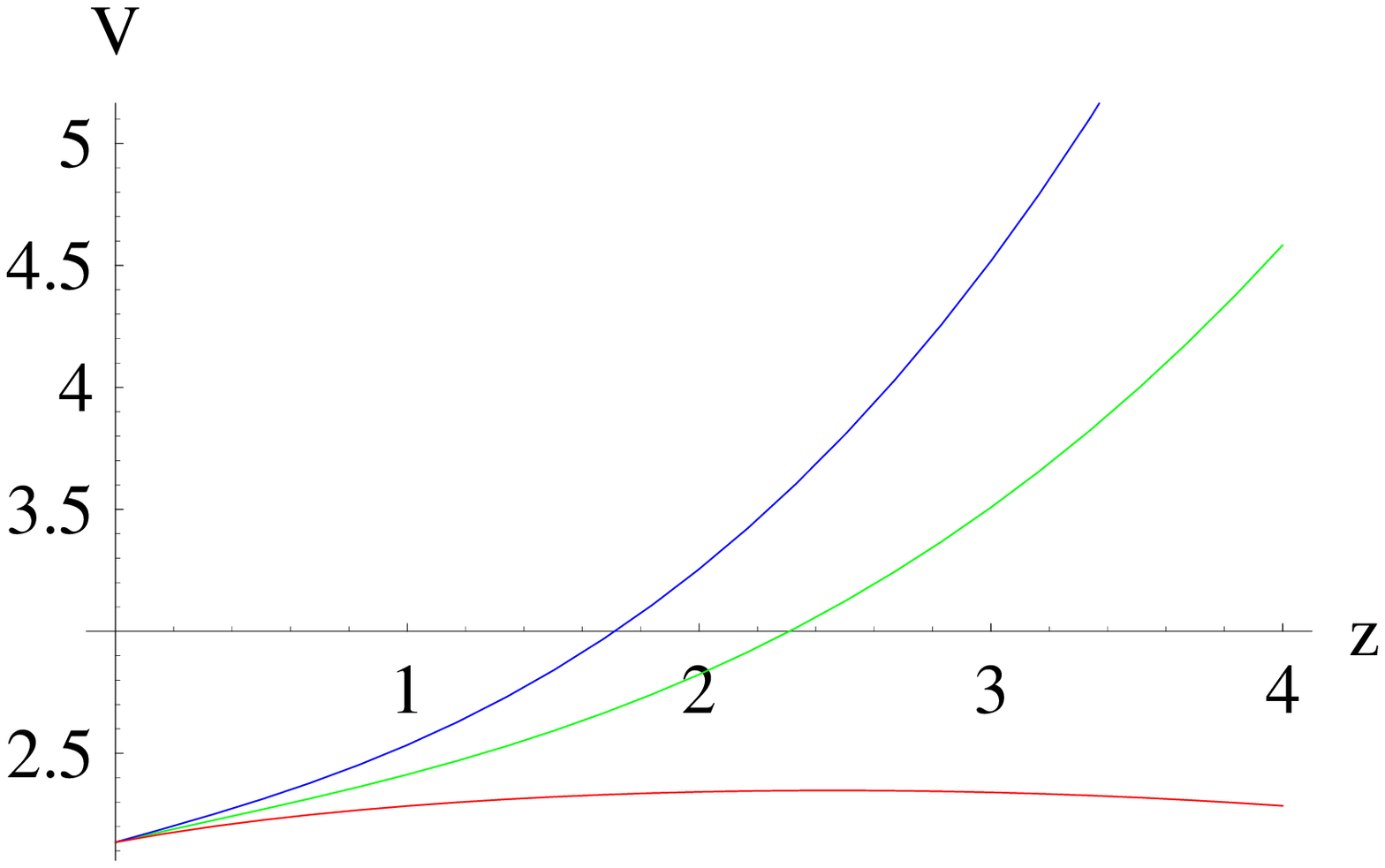}\hskip0.4in
\includegraphics[width=2.8in,height=2.1in]{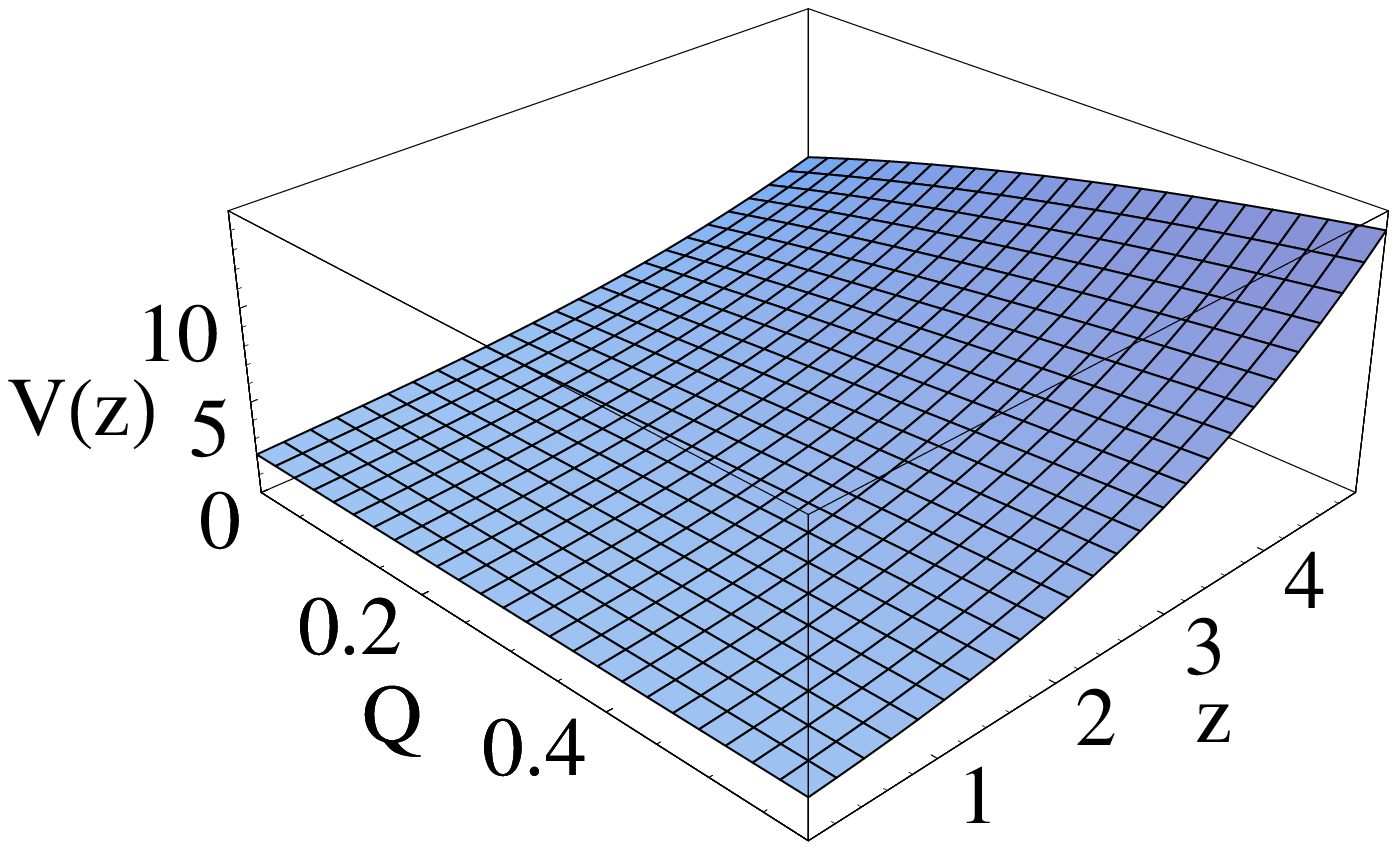}}
\caption{(Left) Form of the scalar potential implied by
(\ref{potentiala}), with $Q=0.2, 0 $ and $-0.2$ (top to bottom)
(blue, green and red). (Right plot) Shape of a reconstructed
scalar field potential.}\label{potential}
\end{figure}

One can think of the scalar field ``rolling down'' the potential,
like a skier on a mountain. It can be seen from
Fig.~\ref{potential} that the slope of the potential now is not
quite flat; rather, the scalar field is rolling with a changing
velocity. This is apparent when taking the time derivative of the
scalar field, $\phi'=-\alpha-2 \zeta e^{2 \zeta N}$. At $N=0$,
$|\phi^\prime|=\alpha+2\zeta$. Indeed, we require that the scalar
field is rolling slowly so that the kinetic energy is small
compared to the potential energy. The scalar field rolls at a
slowly changing velocity for a small value of $\zeta$ and the
acceleration is $\phi^{\prime\prime}(N=0)=-4\zeta^2$, so we expect
that the value of $\zeta^2$ to be much less than $1$. Furthermore,
$\phi^{\prime\prime}$ is negative so the scalar field is slowing
down, it is in a `freezing' phase. The model approaches the
$\Lambda$CDM model as $\alpha, \zeta \to 0$.

The shape of the reconstructed potential shown in Fig.
\ref{potential} is essentially an exponential. In particular, in
the limit $\zeta\to 0$, it is a simple exponential,
$V(\phi)\propto e^{\alpha\phi}$, and for the non-zero $\zeta$ case
this is modified by extra terms proportional to
$e^{c\phi}\,(\phi-\phi_0)^2$. The reconstructed potential shown on
the left panel of Fig. \ref{potential} displays a range of
possible values of $Q$; for larger $Q$ the potential is steeper,
and for smaller $Q$ it is flatter and even changes shape for a
negative $Q$.

\begin{figure}[!ht]
%\FIGURE {\vbox{\vskip 10 pt
\centerline{\includegraphics[width=3.2in,height=2.5in]{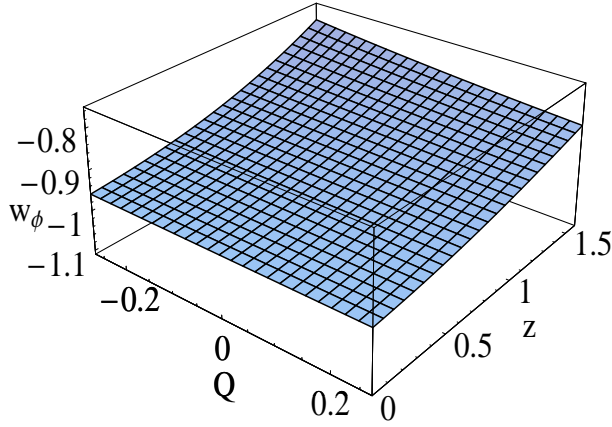}}
\caption{Scalar field equation of state as a function of redshift.
We take the values as $\alpha=0.2$, $\zeta=0.05$ and $\Omega_{0
m}=0.27$.} \label{wde}
\end{figure}

The equation of state for our choice of the parameters is
currently within the limits from WMAP, $-1.11< {\rm w}_\phi <
-0.86$, and does not change much up to redshift $z\sim 1$. The
effect of a nonzero coupling $Q$ on the equation of state is shown
in Fig. \ref{wde}. At low redshift, a positive value of Q drives
${\rm w}_\phi$ closer to $-1$, whereas a negative Q gives a high
value of ${\rm w}_\phi$ at high redshift.

\section{Effect on the background}

\begin{figure}[!ht]
%\FIGURE{\vbox{\vskip10 pt
\centerline{\includegraphics[width=3.0in,height=2.0in]{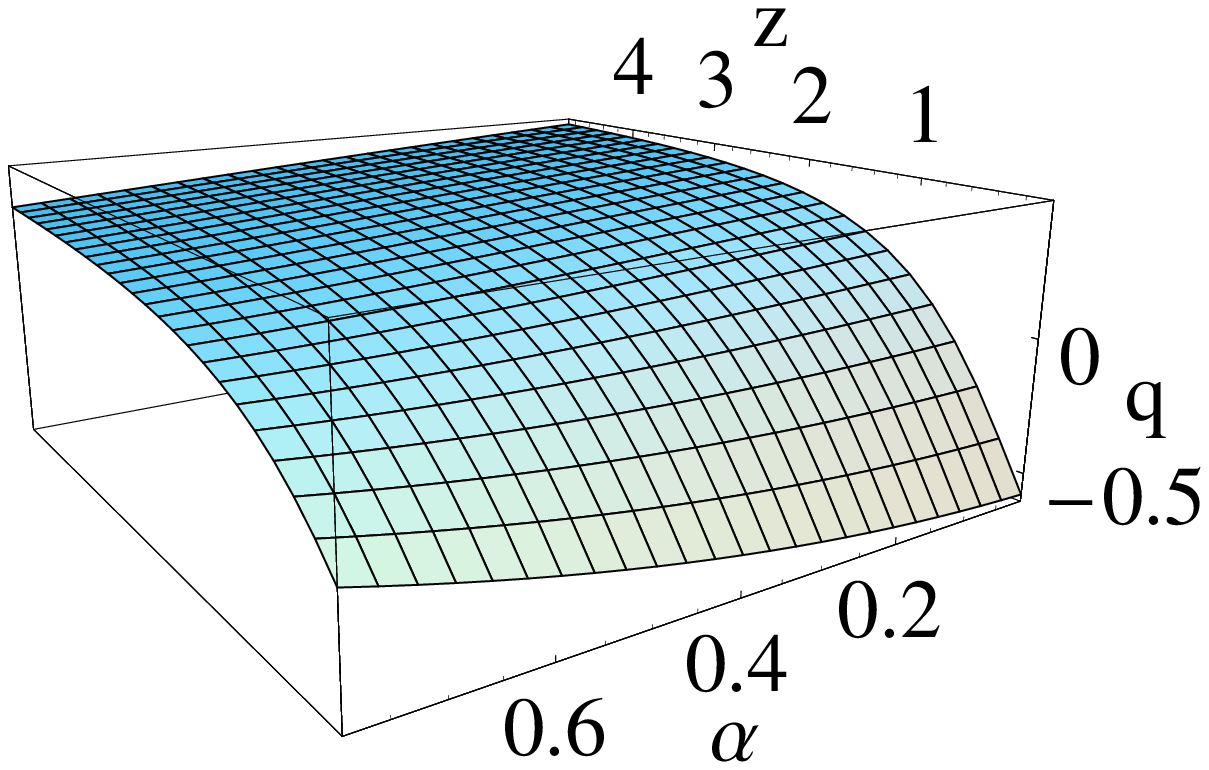}\hskip0.1in
\includegraphics[width=2.7in,height=2.0in]{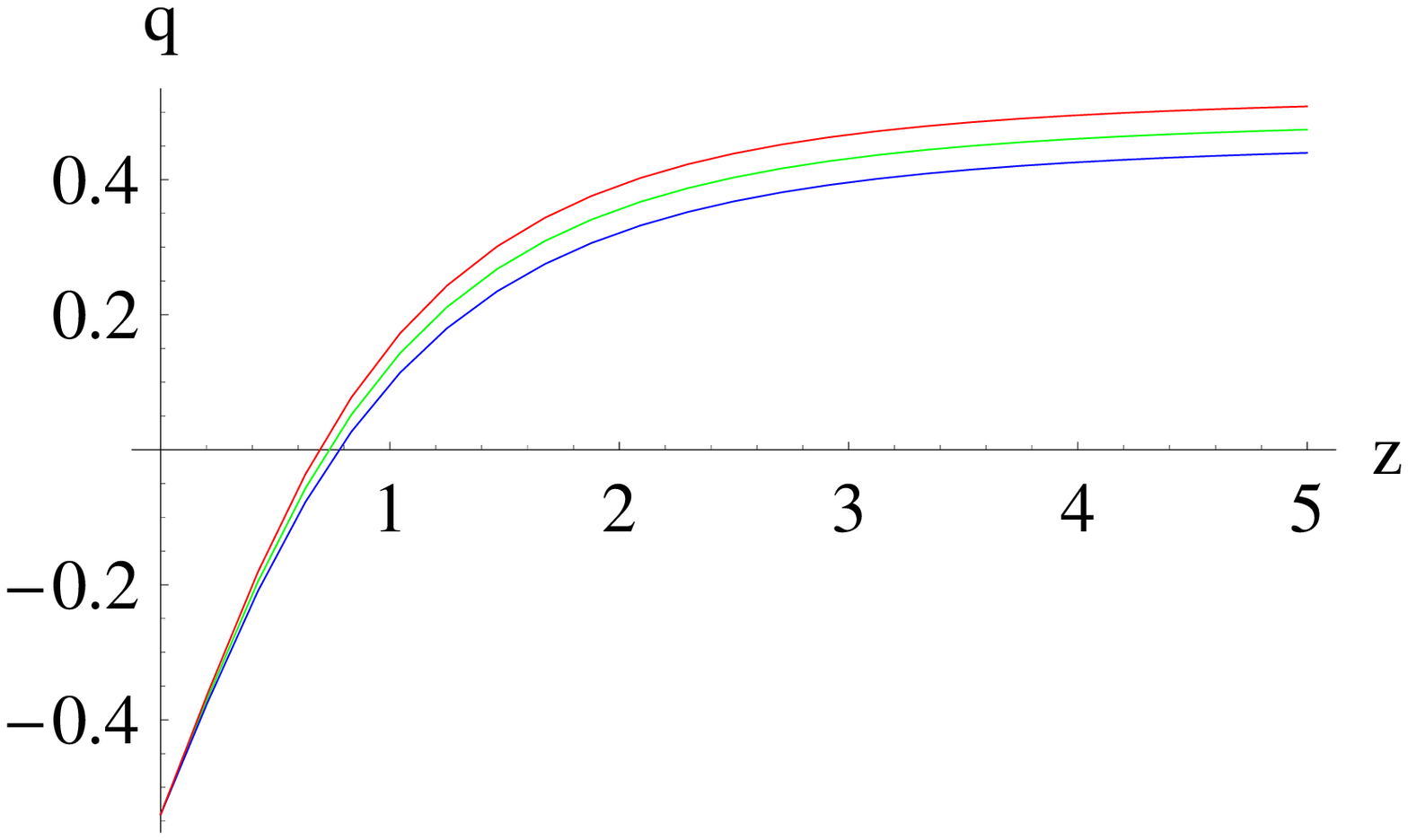}}
\caption{(Left) Deceleration parameter for varying $\alpha$
($\zeta=0.05$, $Q=0.2$). (Right) History of cosmic acceleration
for $Q=0.2, 0$ and $-0.2$ (bottom to top) ($\zeta=0.05$,
$\alpha=0.2$).} \label{qplot}
\end{figure}

An intrinsic property of dark energy is that is does not interact
with light. The only way that it may be observed is through its
effect on the evolution of the background and its possible
interaction with dark matter. In this section we look into how
this model for quintessential dark energy affects the background
by quantitatively looking at its effects on the rate of cosmic
expansion and the fractional densities of dark matter and dark
energy. A useful parameter used when considering cosmic
acceleration is the ``deceleration parameter", $q$. In an
accelerating epoch, $q <0$. In Fig.~\ref{qplot}, we have shown a
typical variation in the value of $q$ with redshift. The important
feature of this reconstruction is that $q$ drops from positive to
negative at $z\approx0.7$, when cosmic acceleration
began~\cite{Melchiorri:2007}. A nonzero coupling has an effect on
the redshift when cosmic acceleration started, $z_{acc}$; the
positive coupling ($Q>0$) implies an earlier start, and the
negative coupling ($Q<0$) implies
a later start. %Within the errors of our model, we have $0.6<z_{\rm
%acc}<0.9$, which is within the predicted redshift for the $\Lambda
%CDM$ model, $z_{acc}=0.76\pm0.10$
%\cite{supernovae,Melchiorri:2007}.

\begin{figure}[!ht]
%\FIGURE{\vbox{\vskip 10 pt
\centerline{\includegraphics[width=2.8in,height=2.1in]{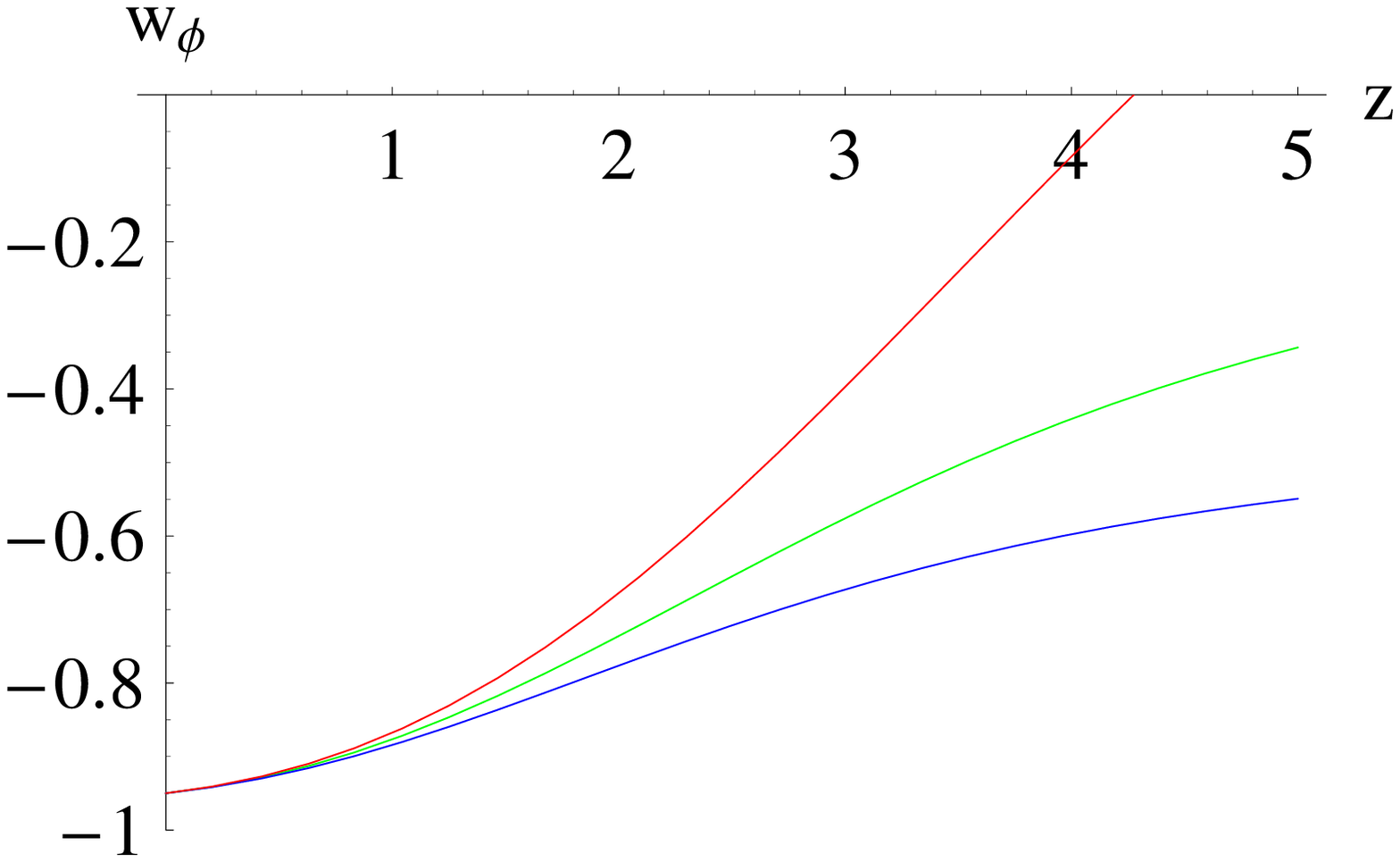}\hskip0.1in
\includegraphics[width=2.8in,height=2.1in]{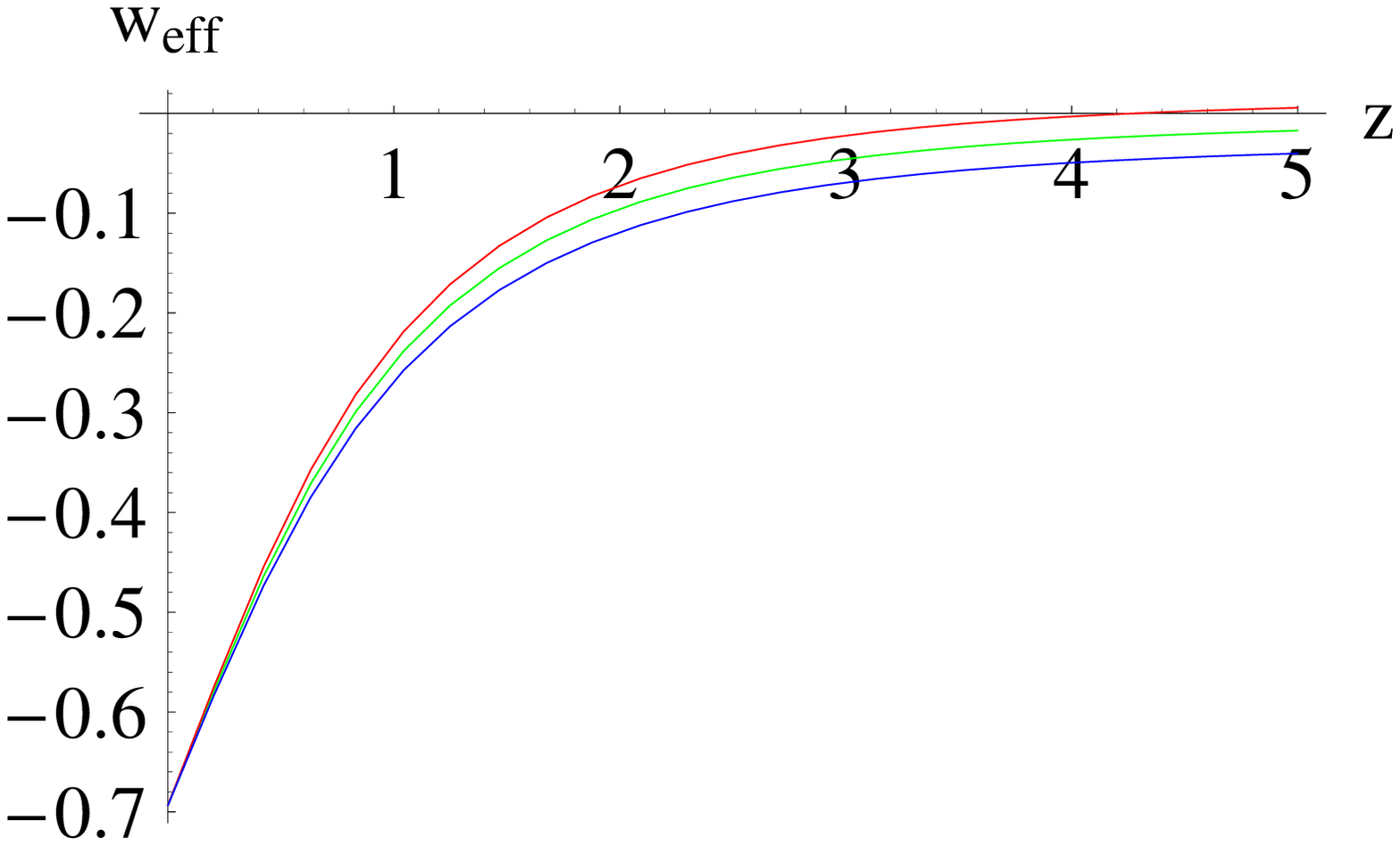}}
\caption{(Left plot) The scalar field equation of state and (right
plot) the effective equation of state with $\zeta=0.05$,
$\alpha=0.2$ and $Q=-0.2, 0$ and $+0.2$ (from top to bottom).}
\label{wdeweff}
\end{figure}

An alternative way of looking at the start of cosmic acceleration
is to see when the effective equation of state drops below $-1/3$
in Fig.~\ref{wdeweff}. At high redshift, the effective equation of
state goes towards $0$, which is expected in the matter-dominated
epoch (${\rm w}_m\simeq0$). For all values of $\alpha$, the
effective equation of state goes to $\sim 0$ during the matter
dominated epoch, but for large values of $\alpha$, ${\rm w}_{\rm
eff}>0 $. This is because the relative energy density of dark
matter and dark energy now is fixed and the matter equation of
state is constant, but the scalar equation of state is closer to
$0$ for a larger $\alpha$, as shown in Fig.~\ref{contour}. At
least, at low redshifts ($z<1$), the effective equation of state
does not vary much for different values of the coupling.

\begin{figure}[!ht]
%\FIGURE{\vbox{\vskip 10 pt
\centerline{\includegraphics[width=2.8in,height=2.0in]{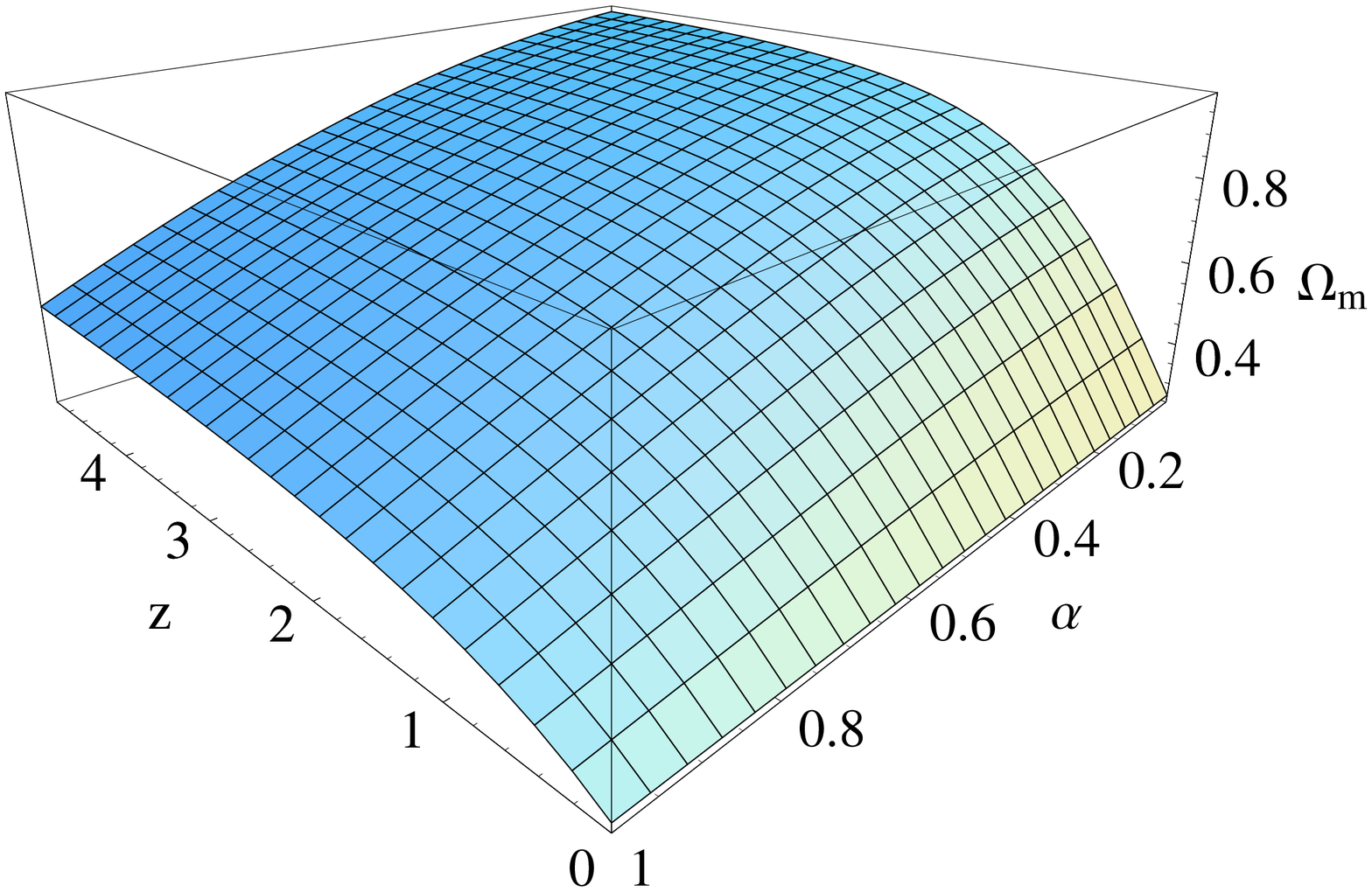}\hskip0.1in
\includegraphics[width=2.8in,height=2.0in]{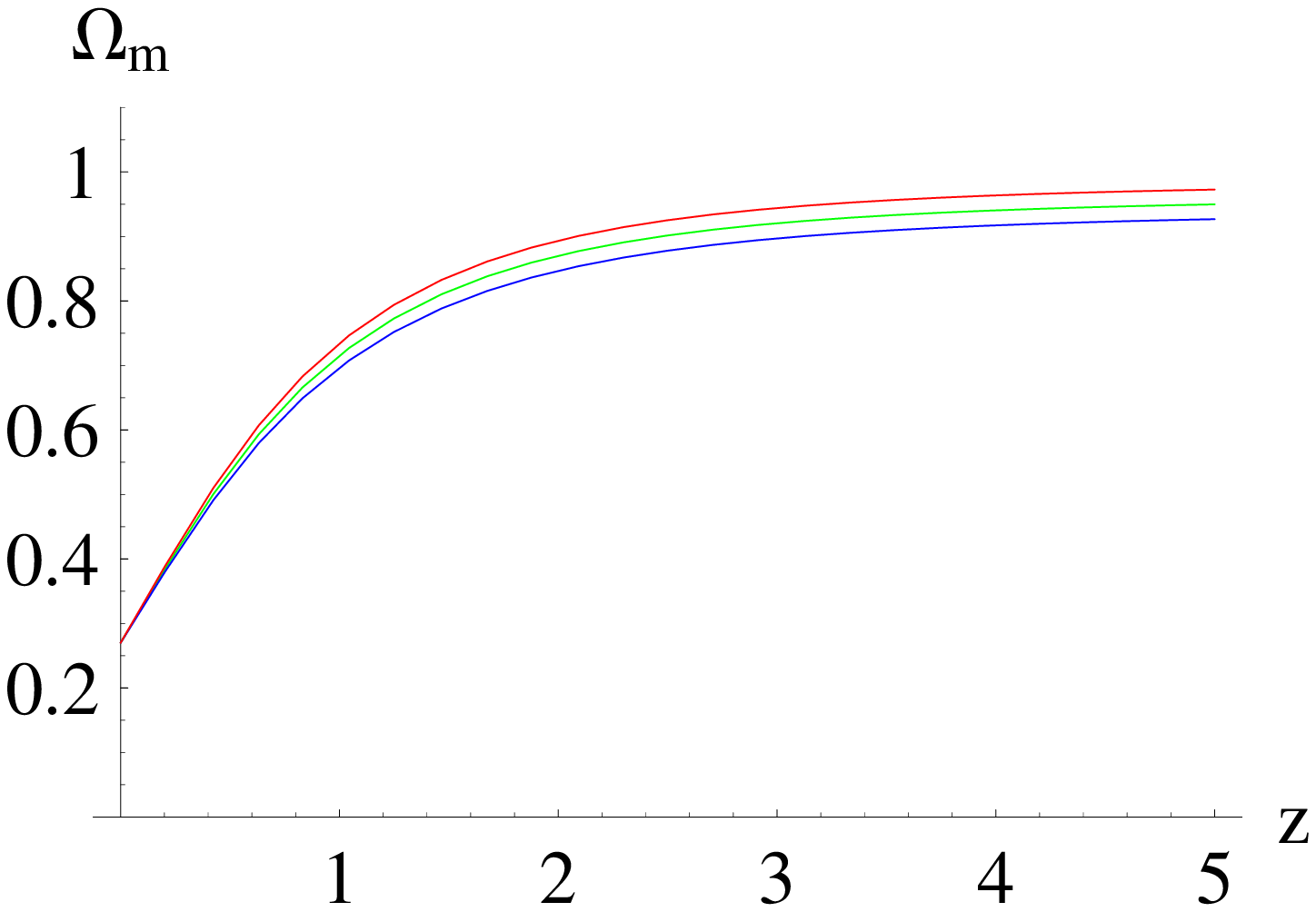}}
\caption{(Left plot) Matter density for varying $\alpha$
($\zeta=0.05$, $Q=0.2$) with $\Omega_{0m}=0.27$. (Right plot)
Matter density for $Q=-0.2, 0$ and $0.2$ (top to bottom; online:
red, green and blue) ($\zeta=0.05$, $\alpha=0.2$).}\label{density}
\end{figure}

The relative energy density of dark matter to dark energy is
affected by changes in the $\alpha$ parameter, as seen in
Fig.~\ref{density}. For smaller values of $\alpha$, the matter
density tends towards a lower value at high redshift. That is, due
to the Friedmann constraint (\ref{vary10}), there is a higher
proportion of dark energy during the matter dominated epoch for a
small $\alpha$ component. Here, it is clear that there has been an
assumption made for the matter density today. The right hand side
plot of Fig.~\ref{density} illustrates the effect of a non-minimal
coupling $Q$ on the evolution of matter density. The effect is not
great within this small range, but there is a difference. A
smaller value of the coupling parameter, $Q$, leads to an increase
in the proportion of matter in the matter-dominated epoch. This
study is valid for a range of low redshift, but will break down at
high redshift when we enter the radiation-dominated epoch because
we have made the assumption that the radiation component is
negligible.

One requirement of the potential is that it should dominate the
kinetic term so that the scalar field equation of state, ${\rm
w}_{\phi}$, may be driven towards $-1$. It can be seen that this
is satisfied by examining Figure~\ref{wdeweff}. Recently, the
authors of~\cite{Sahni:2008xx} have found two new diagnostic tools
for distinguishing quintessential dark energy from Einstein's
cosmological constant. Their method may be extended to a
non-minimally coupled theory for which $|Q|>0$. Here we simply
note that for small values of $\alpha$ and $\zeta$, satisfying the
constraints given in table 2, the constraints of the present model
encompass the cosmological constant limit at the $1\sigma$ level
at low redshifts.

\section{Conclusion}

In this paper, a scalar-tensor theory of quintessential dark
energy has been reconstructed. This reconstruction rests on the
general action (\ref{my action}), in which the tensor theory of
general relativity is modified by including a fundamental scalar
field. The scalar field affects the gravitational part of the
action by introducing a gravitationally repulsive term which
drives the cosmic acceleration, and is coupled to the matter part
of the action, giving rise to a nontrivial dark matter - dark
energy interaction. As for all quintessence models, this action
reduces to the one of general relativity, when the scalar field is
time-independent and the coupling of dark matter to a scalar field
(or dark energy) is minimal.

We made an ansatz for the form of the scalar field itself, in
order to be able to reconstruct relevant cosmological parameters,
including $\Omega_m, \Omega_\phi, \epsilon$ and ${\rm w}_\phi$.
The ansatz $|\phi_0 - \phi|/m\Z{P}= {\alpha} \ln a+ \exp
\left(2\zeta\ln a \right)+{\rm const}$ for the scalar field was
motivated by generic solutions of the dilaton field in some
effective string theory models in four dimensions. The two new
parameters, $\alpha$ and $\zeta$ as well as the matter scalar
coupling parameter, $Q$, were constrained using several
cosmological methods. Primarily, the relationship between $\alpha$
and $\zeta$ was constrained by curve-fitting to various
cosmological data sets. For example, with $\Omega_{0m}=0.27$ and
minimal coupling, using just the SN1a data, minimizing $\chi^2$
gave $\alpha+2 \zeta=\pm 0.44\,(\pm 0.34)$ and using all the data
sets available gave $\alpha+2 \zeta=\pm 0.3\,(\pm 0.3)$.
%The value of the coupling was loosely constrained to $Q=-0.7\pm1.1$
%when fitted to the data along with $\alpha+2\zeta=0.0\pm3.8$.

We also considered the effect of dark energy on the growth of
large scale structure. The effect of a nonminimal coupling on the
standard expression for the rate of growth of matter
perturbations, $f=\left[\Omega_m(N)\right]^\eta$, was considered.
A modified ansatz for the dependence of matter density and dark
matter-dark energy coupling on the matter fluctuation growth rate
was introduced and then fitted to some observed values of the
growth rate at low redshift. This analysis gave a best-fit minimum
value for the coupling $Q=0.2\,\pm 0.2$ (with the input $\Omega_{0
m}\simeq 0.27$), which is compatible with the minimally coupled
scalar field case. These methods of constraining the variables of
the model have succeeded in constraining $\alpha+2\zeta$, but
failed in constraining the separate values of $\alpha$ and
$\zeta$. This is a direct result of taking the small $\zeta$ limit
when analyzing the solution for $\Omega_m$. Nevertheless, by
assuming the same field for the inflaton as quintessence the
individual values of $\alpha$ and $\zeta$ were constrained using
the spectral index of the CMB. We find $\alpha=0.20\pm0.04$ and
$\zeta=0.04\pm0.15$.

Our analysis showed that the present model of coupled quintessence
is compatible with a cosmological constant (where $\alpha, \zeta =
0$ and $Q=0$), within $1\sigma$ error bars. This means that dark
energy may still simply be a cosmological constant, but even a
small deviation away from this limit results in a very different
source of cosmic acceleration, with time varying behavior. The
number of parameters of this model may make it hard to rule out,
as they may be tuned to satisfy many more constraints. In this
work, the matter scalar coupling parameter was chosen to be a
constant. This was not motivated physically but was chosen for the
solvability of the system. Indeed, there is no physical reason why
the dark energy - (dark) matter interaction should not vary in
time, it is in all likelihood a function of $\phi$. There needs to
be further study to explore this $\phi$ dependence of $Q$. Also,
the links between quintessential dark energy and inflation deserve
further investigation.

% Bibliography

%\bibliography{DE}
%\bibliographystyle{plain}

\end{document}